\documentclass[a4paper, 11pt]{article}
\pdfoutput=1
\usepackage{jcappub}
\usepackage[T1]{fontenc}
\usepackage{subfigure}
\usepackage{natbib}
\usepackage{txfonts,makecell}   
\usepackage{hhline}
\usepackage{hyperref}
\usepackage{cleveref}
\usepackage{array}
\usepackage{amsmath}
\usepackage{soul}
\usepackage[numbers]{natbib}
\crefname{equation}{Eq.}{Eqs.}
\newcounter{mysfig}
\renewcommand\themysfig{(\alph{mysfig})}
\makeatletter
\newcommand\Scaption[1]{%
\refstepcounter{mysfig}%
\vskip.5\abovecaptionskip
  \sbox\@tempboxa{\small\themysfig~#1}%
  \ifdim \wd\@tempboxa >\hsize
    \small\themysfig~#1\par
  \else
    \global \@minipagefalse
    \hb@xt@\hsize{\hfil\box\@tempboxa\hfil}%
  \fi
  \vskip\belowcaptionskip}
\makeatother

\newcommand*\aap{Astron.~Astrophys.}

\newcommand*\ao{Appl.~Opt.}

\newcommand*\apjl{Astrophys.~J.~Lett}

\newcommand*\apjs{Astrophys.~J.~S}

\usepackage{hyperref}
\usepackage{cleveref}
\usepackage[dvipsnames]{xcolor}
\usepackage{booktabs}

\newcommand{\dr}[1] {\frac{\text{d}#1}{\text{d}r}}

\newcommand{\Mpc}{\text{Mpc}}
\newcommand{\Mp}{M_{\rm Pl}}
\newcommand{\Veff}{V_{\rm eff}}

\newcommand{\rc}{r_{\rm c}}
\newcommand{\xc}{x_{\rm c}}

\newcommand{\Qtwo}{\ensuremath{\mathcal{Q}_{2}}}
\newcommand{\phitwo}{\ensuremath{\phi_{2}}}
\newcommand{\rhos}{\ensuremath{\rho_{\rm s}}}
\newcommand{\rs}{\ensuremath{r_{\rm s}}}
\newcommand{\Cs}{\ensuremath{C_{\rm s}}}

\newcommand{\logB}{\log(\mathcal{B}) }
\newcommand{\phiinf}{\ensuremath{\phi_{\infty}}}
\newcommand{\phiB}{\frac{\phi_\infty}{\mathcal{B}}}

\title{Mass Modeling and Kinematics of Galaxy Clusters in Modified Gravity}

\author[a]{Lorenzo Pizzuti,}
\author[b,c,d,h]{Yacer Boumechta,}
\author[b,c,d]{Sandeep Haridasu,}
\author[f,g]{Alexandre M. Pombo,}
\author[a]{Sofia Dossena,}
\author[b,c,d]{Minahil Adil Butt,}
\author[b,c,d]{Francesco Benetti,}
\author[b,c,d]{Carlo Baccigalupi,}
\author[b,c,d,e]{Andrea Lapi}

\affiliation[a]{Dipartimento di Fisica G. Occhialini, Universit\'a degli Studi di Milano Bicocca, Piazza della Scienza 3, I-20126 Milano, Italy}
\affiliation[b]{SISSA, Via Bonomea 265, 34136 Trieste, Italy}
\affiliation[c]{Institute for Fundamental Physics of the Universe (IFPU), Via Beirut 2, 34014 Trieste, Italy}
\affiliation[d]{INFN-Sezione di Trieste, via Valerio 2, 34127 Trieste, Italy}
\affiliation[e]{IRA-INAF, Via Gobetti 101, 40129 Bologna, Italy}
\affiliation[f]{CEICO, Institute of Physics of the Czech Academy of Sciences, Na Slovance 2, 182 21 Praha 8, Czechia}
\affiliation[g]{Dipartimento di Fisica G. Occhialini, Universit\'a degli Studi di Milano Bicocca, Piazza della Scienza 3, I-20126 Milano, Italy}
\affiliation[h]{ICTP-The Abdus Salam International Centre for Theoretical Physics, Strada Costiera 11, 34151 Trieste, Italy  }

\emailAdd{lorenzo.pizzuti@unimib.it, yboumech@sissa.it, sharidas@sissa.it, pombo@fzu.cz}

\abstract{The chameleon screening mechanism has been constrained many a time using dynamic and kinematic galaxy cluster observables. Current constraints are, however, insensitive to different mass components within galaxy clusters and have been mainly focused on a single mass density profile, the Navarro-Frenk-White mass density model. In this work, we extend the study of the Chameleon screening mechanism in galaxy clusters by considering a series of mass density models, namely: generalized-Navarro-Frenk-While, b-Navarro-Frenk-While, Burket, Isothermal and Einasto. The coupling strength ($\beta$) and asymptotic value of the chameleon field ($\phi_\infty$) are constrained by using kinematics analyses of simulated galaxy clusters, generated both assuming General Relativity and a strong chameleon scenario.  By implementing a Bayesian analysis we comprehensively show that the biases introduced due to an incorrect assumption of the mass model are minimal. Similarly, we also demonstrate that a spurious detection of evidence for modifications to gravity is highly unlikely when utilizing the kinematics of galaxy clusters. }

\keywords{Cosmology -- Galaxy clusters -- Modified gravity}

\begin{document}
\maketitle
\flushbottom


%
\section{Introduction}\label{sec:Intro}
%
    With the advent of high-accuracy cosmological observations, the possibility to exclude or further consider models that are able to describe the observable universe across its cosmological scales was open. Together with a cosmological constant ($\Lambda$)~\cite{Weinberg:1988cp, Bull:2015stt}, General Relativity (GR) is able to explain, with a high degree of precision, most cosmological observations \cite{Adam:2015rua, Ade:2015xua}. While the latter describes the short-range, strong field regime, the former accounts for the late-time acceleration of the Universe \cite{Riess:1998cb,Ravi:2023nsn,Ravi:2024jwl,Solanki:2022rwu}, which is now a well-established phenomenon \cite{Planck:2015lwi, haridasu2017strong}.

    However, an incompatibility between the theoretical~\cite{rugh2009physical,SolaPeracaula:2022hpd,Carroll:2000fy} and the observational value of cosmological constant exists, creating what has been called "the worst theoretical prediction in the history of physics"~\cite{Weinberg:1988cp, carroll1992cosmological, dechant2009anisotropic}. Despite its successes (\textit{e.g.} \cite{Planck2020}), the $\Lambda$-cold dark matter model ($\Lambda$CDM) still fails in providing a natural explanation for the existence and physical origin of the cosmological constant\cite{Bernardo:2022cck}. In this regard, several attempts have been proposed in the last decades, of particular focus are modes that consider additional degrees of freedom (\textit{e.g.} quintessence \cite{Caldwell:1997ii, Caldwell05, Linder02}) and modifications of GR \cite{Nojiri:2017ncd, Shankaranarayanan:2022wbx, Clifton:2011jh}. One of the most popular classes of modified theories of gravity occurs when adding a scalar field (\textit{aka} additional degree of freedom) to the original Lagrangian. These are known as scalar-tensor theories\cite{Moffat:2005si,SolaPeracaula:2019zsl}. The presence of such a scalar field provides a further contribution to the gravitational force~\cite{Khoury:2013tda, Burrage:2017shh}, leaving detectable imprints on the formation and evolution of cosmic structures \cite{Brax:2005ew, Brax:2004qh, Brax:2013mua}. However, while at large scales, observations are compatible with the effect of the additional scalar field, in high-density environments\cite{Lombriser:2013eza,Boriero:2015loa}, simple GR has been consistently proven to be the best description of the universe. A screening mechanism is then required in order to suppress this new interaction on small scales. Depending on the implementation of this screening mechanism, the latest (fifth) force on matter density perturbations can be significantly different\cite{Briddon:2023ayq,Pernot-Borras:2019gqs}.

    Among the plethora of possibilities, Chameleon field theories \cite{Khoury:2003aq, Faulkner:2006ub, Navarro:2006mw} deliver some of the most promising results. In the latter, the environment-dependent effects of gravity are achieved by a non-minimally coupling between an additional scalar field (\textit{aka} chameleon field) and matter, with the screening achieved by the scalar fields' potential dependency on the local energy density\cite{Waterhouse:2006wv}. The resulting effective mass becomes very large in high-density regions (suppressing the interaction) while tending to a small (non-zero) value in low-density regions, resulting in an effective fifth force \cite{Khoury:2003aq} that mimics the influence of a positive cosmological constant in the motion of non-relativistic objects. The scalar field is usually characterized by two parameters: the coupling constant between the scalar field and matter, $\beta$, and the value of the field at infinity, $\phi _\infty$, where the matter density matches the background density -- the maximum value of the Chameleon field -- under reasonable assumptions (\textit{e.g.} \cite{Terukina:2013eqa}), these parameters describe entirely the modification of gravity. Due to a large amount of high-accuracy observational data, Chameleon theory parameters are tightly constrained, at the laboratory (\textit{e.g.} \cite{Burrage_2015, pernotborras:insu-03182938}), astrophysical (\textit{e.g.} \cite{Naik_2018, Dima_2021, Desmond2020,Benisty2023}) and cosmological scales (\textit{e.g.} \cite{Wilcox:2015kna, Cataneo:2016iav, Pizzuti17, Tamosiunas_2022}), see also \cite{Burrage:2017shh} for a review. Although the region of the parameter space for viable Chameleon theories has been tightly constrained, there is yet a significant leeway in galaxy cluster scales. The majority of the Chameleon field constraints with clusters were obtained by assuming only the Navarro-Frenk-White (NFW) profile~\cite{Navarro:1995iw} to model the entirety of the cluster's mass distribution (baryonic+dark matter).

    In our previous work \cite{Boumechta:2023qhd}, based on an analysis already performed in \cite{Ettori:2018tus}, the NFW mass density profile was assumed to describe the total mass density distribution of a compilation of galaxy clusters\cite{Ettori:2018tus}. Nevertheless, this assumption was valid in the GR context only for 9 of the clusters. For the rest of them, the NFW mass density profile is not the preferred model, for instance, favouring an Isothermal or a Hernquist mass profile. This preference for a different mass mode other than NFW can yield a strong effect on the Chameleon field profile -- and resulting fifth force -- with the shape of the matter density. Allowing one to test the rich phenomenology of the screening mechanism at galaxy cluster scales for additional mass density models is tantamount.
    
    The aim of the current manuscript is then to investigate the effect of the mass modelling when constraining Chameleon gravity at cluster scales by current mass measurements. In order to test the viability and effect of several mass density models on the Chameleon screening, we first implement a semi-analytical approximation to obtain the solution of the Chameleon field equation for six different mass model assumptions. These are then utilised in simulating reliable kinematic data of cluster member galaxies complemented by lensing-like information. We then assess the biases introduced due to in-congruent assumptions of mass models in simulated data and the likelihood analysis. The semi-analytical approximation provides a rather simple and computationally cost-efficient way to implement the screening mechanism for operative purposes, with respect to the full numerical solution. Moreover, it is a powerful tool to study the relationship among the parameters of the mass profile and of the Chameleon field, highlighting the main physical properties of the screening mechanism. To validate our semi-analytical approach we also perform a comparison with a full numerical solution, successfully confirming  that the former can adequately reproduce the behavior of the fifth force.

    The paper is structured as follows: \cref{sec:Theory} reviews the basic theory of the Chameleon gravity and the screening mechanism, further presenting the mass models adopted and the semi-analytical approximation in \cref{sec:mass_models} and \cref{sec:field_solutions}, respectively. In \cref{sec:solutions}, the solutions for the field profile and its derivative are displayed and tested against the numerical solutions in \cref{sec:numerical}. \cref{sec:kinematics} briefly describes the simulation of synthetic clusters and the \textsc{MG-MAMPOSSt} code \cite{Pizzuti:2022ynt} utilized for the kinematic mass profile reconstruction; the results of the analysis are then presented and discussed. Finally, \cref{sec:conclusions} summarizes the main conclusions and the key findings. 

%
\section{Theory}\label{sec:Theory}
%
    In this work, the Chameleon field theory~\cite{Khoury:2003aq}, and respective phenomenological effect, will be studied in the context of galaxy clusters obeying the assumptions: \textit{i)} the cluster possesses radial symmetry\footnote{Please see \cite{Tamosiunas_2022} for a discussion generalising the solutions with triaxiality.}, and \textit{ii)} both dark matter and baryonic matter are modelled with the same total mass profile~\cite{Boumechta:2023qhd}. Let us start by describing the Chameleon mechanism.
%
    \subsection{Lagrangian and the equation of motion}
%
    In Chameleon gravity, a real scalar field, $\phi$ is conformally coupled to the matter fields $\psi^{(i)}$, and the Lagrangian~\cite{Khoury:2003aq, Faulkner:2006ub} comes as
        \begin{equation} \label{eq:lagrangian}
         \mathcal{L}=\sqrt{-g}\left\{ -\frac{\Mp R}{2}+\frac{(\partial\phi)^{2}}{2}+V(\phi)\right\} +L_{\rm m}\big( \psi^{(i)},g_{\mu\nu}^{(i)}\big)\ ,
        \end{equation}
    with $\Mp =1/\sqrt{8\pi G}$ the reduced Planck mass, while $g_{\mu\nu}^{(i)}= e^{-\frac{2\beta_{i}\phi}{c^2 \Mp}} \tilde{g}_{\mu\nu}$, with $g_{\mu\nu}^{(i)}$ the metric in the Einstein frame and $ \tilde{g}_{\mu\nu}$ the metric in the Jordan frame; $\beta_i$ the coupling strength of the Chameleon field to each matter field, $\psi ^{(i)}$, and $c$ is the speed of light. Following the standard procedure, for simplicity one considers a single effective matter field, $\psi$, that describes the entire matter distribution\footnote{While it is possible to extend the analysis to distinct field couplings to the chameleon field, it falls away from the objective of the manuscript. It will be explored in future investigations.}. This results in a single coupling constant, $\beta$\footnote{A particular choice for the coupling, $\beta = 1/\sqrt{6}$, corresponds to the popular class of $f(\mathcal{R})$ gravity models \cite{starobinsky2007disappearing, oyaizu2008nonlinear}.}, and Einstein frame metric $g_{\mu\nu}=e^{-\frac{2\beta \phi}{c^2 \Mp}} \tilde{g}_{\mu\nu}$.
    
    Non-relativistic matter have an energy density $\tilde{\rho}=\rho\, e^{\frac{\beta\phi}{c^2 \Mp}}$. The potential $V(\phi)$ is a monotonic function of the scalar field.
    
    The resulting equation of motion for a static Chameleon field is then,
        \begin{equation} \label{eq:eq_motion}
         \nabla^{2}\phi=\frac{\partial}{\partial\phi}\Veff(\phi)\ ,
        \end{equation}
    where the effective potential is defined as 
        \begin{equation}\label{eq:effective_potential}
         \Veff(\phi)=V(\phi)+\rho\, e^{\frac{\beta\phi}{c^2 \Mp}} .
        \end{equation}
    A typical form of the potential assumes a power-law $V(\phi) =\Lambda^{4+n}\phi^{-n}$ where $n$ and $\Lambda$ are constants that define the model \cite{Terukina12, Terukina:2013eqa}. Note that $\phi/\Mp$ has the dimension of energy per unit mass, and $\Lambda$ is an energy scale commonly set to the dark energy value (\textit{e.g.} \cite{Tamosiunas_2022}). In the limit $\beta\phi/(c^2 \Mp)\ll1$, which is valid throughout this context given current constraints on the field, the effective potential can be approximated by $\Veff(\phi)\simeq V(\phi)+\rho\, \big( 1+\beta\phi/(c^2 \Mp)\big)$. The resulting equation of motion \eqref{eq:eq_motion} is, 
        \begin{equation} \label{eq:eq_of_motion_s}
         \nabla^{2}\phi=\frac{\beta}{c^2 \Mp}\rho+V'(\phi)\ .
        \end{equation}
    Solving \cref{eq:eq_of_motion_s} under appropriate boundary conditions -- namely, in the absence of a matter-density distribution, the scalar field tends to a constant, background, value, $\phi (r\to \infty) =\phi _B$; while at the centre of the system, the matter-density distribution is constant and the scalar field remains suppressed, $\nabla ^2 \phi \approx 0$  -- for a given mass distribution yields the profile of the Chameleon field, which in turn, originates the fifth force. 
    
   In order to describe the mass distribution of a galaxy cluster, one has to resort to a mass density profile, $\rho$. So far, the majority of the studies of Chameleon gravity at cluster scales in the literature relied on the standard NFW density model. However, not all galaxy clusters are best described by the latter\footnote{Bayesian analysis done in \cite{Ettori:2018tus}, suggests that the three clusters in the GR case prefer either Isothermal and Burket profiles, albeit a weak preference, with the change in Bayesian evidence $\lesssim 2.5$.}. It is then tantamount to extend the analysis to alternative mass density models, and explore the respective Chameleon screening phenomenology in order to assess its constraints and thus investigate whether such mass models could be validated in the context of Chameleon gravity or by any means change the evidence.
%
    \subsection{Mass models} \label{sec:mass_models}
%
    As stated, while the NFW mass density model is able to provide an adequate fit to the total mass distribution of galaxy clusters, some discrepancies have been found, both in simulations and observations, concerning the inner shape of cluster-size halos (\textit{e.g.} \cite{Sereno_2016,Peirani17}) especially when baryonic feedback is considered. Moreover, it is not completely clear if the NFW mass density model is a good description of halos in modified gravity scenarios (see \textit{e.g.} \cite{Corasaniti20} and references therein), even if some works indicate that it performs well in reproducing the mass distribution in chameleon gravity (\textit{e.g.} \cite{Lomb12}).
    
    In this work, besides the NFW profile, five additional mass-density models will be considered. These are: $b$NFW \cite{Terukina12}, generalised NFW \cite{Zhao96} ($g$NFW), Burkert \cite{Burkert:2000di}, Isothermal \cite{King1962} and Einasto \cite{Einasto1965}. All these mass density models are characterized by a central density, $\rhos$, and a scale radius, $\rs$, which vary between halos of different sizes. To solve the field \cref{eq:eq_of_motion_s}, we apply and generalise the same method as earlier implemented in \cite{Terukina:2013eqa, Tamosiunas:2021kth, Boumechta:2023qhd, Pizzuti:2020tdl, Pizzuti:2022ynt}, which can be summarized as follows. 
    
    Assuming radial symmetry, the procedure relies on a semi-analytical approximation (see \cite{Terukina12}) by solving \cref{eq:eq_of_motion_s} in the outskirts of the galaxy cluster where the Chameleon field becomes important (low-density regime). The Chameleon field is assumed to be negligible in the innermost region, where the field is screened (high-density regime). The transitional screening radius between the two regions, $\rc$, can be determined by imposing the continuity of $\phi(r)$ and its first derivative at $\rc$ (\textit{aka} junction condition). The resulting screening equation relates the parameters of the Chameleon model ($\beta$ and $\phi _\infty$) and that of the mass model ($\rhos$ and $\rs$).

    The aforementioned NFW profile is described by,  
        \begin{equation} \label{eq:NFW}
         \rho(r)=\frac{\rhos}{\frac{r}{\rs}\left( 1+\frac{r}{\rs}\right)^2}\ ,
        \end{equation}
   which, while simple, has been extensively used to describe the mass distribution of galaxy clusters and constrain the parameters of the Chameleon field.

   A straight forward extension of the NFW profile -- ($b$NFW~\cite{Terukina12}) -- is obtained by considering a generic integer exponent $b>2$ in the denominator of \cref{eq:NFW}\footnote{For $b=3$ this is the so-called Hernquist profile \cite{Hernquist1990}.}:

        \begin{equation} \label{eq:bNFW}
         \rho(r)=\frac{\rhos}{\left(\frac{r}{\rs}\right)\left( 1+\frac{r}{\rs}\right)^{b}}\ .   
        \end{equation}

    Another, more advanced, generalization of the NFW model is the \textit{generalised} NFW ($g$NFW) profile, which is characterized by a real slope $0<\gamma <2$ as 
        \begin{equation} \label{eq:gNFW}
         \rho(r)=\frac{\rhos}{\left(\frac{r}{\rs}\right)^\gamma \left( 1+\frac{r}{\rs}\right)^{3-\gamma}}\ .   
        \end{equation}
   The second distinct mass density profile under consideration is the so-called Burket profile~\cite{Burkert:2000di},
        \begin{equation}\label{eq:bur}
         \rho(r)=\frac{\rhos}{\left( 1+\frac{r}{\rs}\right)\left[1+\Big(\frac{r}{\rs}\Big)^2\right]}\ .
        \end{equation}
    The Isothermal mass density model~\cite{Ettori:2018tus,King1962} is described by
        \begin{equation}\label{eq:Iso}
         \rho(r)=\frac{\rhos}{\Big[ \left(\frac{r}{\rs}\right)^{2}+1\Big]^{3/2}}\ .
        \end{equation}
    Finally, the Einasto model~\cite{Einasto1965} comes as,
        \begin{equation}\label{eq:Ein}
         \rho(r)=\rhos\exp \left \{ -2 m \left[ \Big(\frac{r}{\rs}\Big)^{\frac{1}{m}}-1\right]\right\}\ ,
        \end{equation}
    where $m \in \mathbb{N} $ is a characteristic exponent.

\begin{figure}
         \centering
         \includegraphics[width =1.0\textwidth]{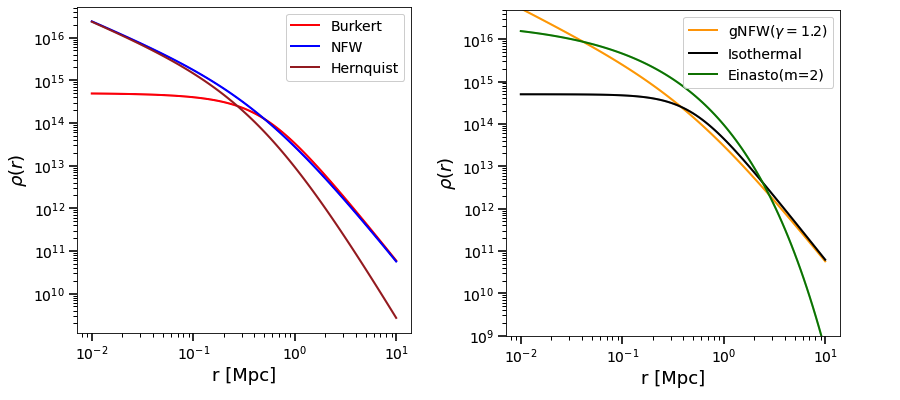}
         \caption{Density profile models used in this work, for a halo with $\rhos = 5\times 10^{14} M_\odot/\text{Mpc}^3 $ and $\rs = 0.5 \,\text{Mpc}$. Left: Burkert, NFW, Hernquist (b-NFW with $b=3$); Right: gNFW with $\gamma = 1.2$, Isothermal and Einasto with $m=2$.}
         \label{fig:densities}
        \end{figure}

    In Fig.~\ref{fig:densities} the density profiles for all the models introduced above are shown for a reference value of the critical density $\rhos = 5\times 10^{14} M_\odot/\text{Mpc}^3 $ and a scale radius $\rs = 0.5 \,\text{Mpc}$. The Isothermal, Burkert and Einasto profiles flatten to a constant value for $r\to 0$, while the NFW and its generalizations exhibit a cusp, diverging at small radii. Increasing $b$ in the b-NFW models provides a faster suppression of the density at large $r$, whereas an increase of the $m$ parameter in the Einasto model results in a shallower profile. In total, the set of density models captures a quite broad range of behaviours.
    While there are additional assumptions that one could utilise for the mass density profiles of galaxy clusters, we find the above selection to be sufficiently extensive for the current analysis: test the effect of the Chameleon mechanism on the shape of the mass distribution.
%
    \subsection{Field solutions}\label{sec:field_solutions}
%
    As a recall, \cref{eq:eq_of_motion_s} is solved by considering two regions. Deep within the massive source -- the centre of the galaxy cluster (high-densities) --, the scalar field is everywhere close to its minimum value, and field gradients are negligible $\nabla^2 \phi \approx 0$ (see \textit{e.g.} \cite{Khoury:2003aq} for details). Thus, from \cref{eq:eq_of_motion_s} and assuming the power-law potential, the scalar field inside the source can be approximated as,
        \begin{equation}\label{eq:inner}
         \phi_\text{int}(r) \approx \left(\beta\, \frac{\rho(r)}{n\Lambda^{4+n}\Mp} \right)^{-\frac{1}{n+1}}\ ,
        \end{equation}
    {where the absence of gradient at the centre of the mass distribution effectively screens the fifth force (see below), assuming $\phi_\text{int}(r) \simeq 0 $ and hence $\phi_\text{int}(r) \ll \phi_\infty$ in the following.} On the other hand, towards the outskirts of the source, the gradient of the field grows and leads to a fifth force given by 
        \begin{align}
         F_{\phi} = -\frac{\beta }{\Mp} \frac{\text{d}\phi}{\text{d}r}\ . 
        \end{align}
    Wherein the above expression corresponds to a faraway region from the massive body -- low-density regime -- and the Laplacian term dominates over the field's potential, which decreases quickly (\textit{i.e.} $\partial V(\phi)/\partial \phi \ll \nabla^2\phi$). The equation of motion for the Chameleon field in this region can then be expressed as,
        \begin{equation}\label{eq:outer}
         \nabla^2 \phi_\text{ext} \approx \beta\frac{ \rho}{\Mp}\ .
        \end{equation}
   {Assuming that the matter density model can be written as $\rho(r)=\rhos\,f(x)$, where $x=r/\rs$, which is valid for all the mass models presented above, \cref{eq:outer} can be expressed as}
        \begin{equation}\label{eq:outeradial}
         x^2\frac{\text{d} \phi_\text{ext}}{\text{d} x}=\frac{\beta \rs^2\rhos}{\Mp}\int{f(x)\, x^2\, \text{d}x}+\Cs\ ,
        \end{equation}

    where $\Cs$ is an integration constant. We now distinguish two cases: $a)$ when the field in the interior region is screened (the aim of this work) and $b)$ when it does not reach the minimum of the effective potential (\textit{i.e.} no interior solution). Case $b)$ is equivalent to a zero screening radius -- or negative -- thus, $\phi(r)$ will not be screened, and the 'exterior' solution is valid everywhere. The constant of integration in this scenario can be determined by fixing the boundary conditions. 
    
    The gravitational potential, $\Phi$, in Chameleon gravity, comes as,
        \begin{align}
         \frac{\text{d}\Phi}{\text{d}r} & = \frac{GM(r)} {r^2}+\frac{\beta}{\Mp}\frac{\text{d}\phi}{\text{d}r}=\frac{G}{r^2}\left[M(r)+\frac{\beta}{G\,\Mp}\, r^2\, \frac{\text{d}\phi}{\text{d}r}\right] \nonumber \\
         & \equiv \frac{G}{r^2}\left[M(r)+M_\text{eff}(r)\right]\ , \label{eq:pot_chameleon}
        \end{align}
    \textit{i.e.}, up to a constant, the field gradient times $r^2$ acts as an additional effective mass contribution sourced by the fifth force. As such, one can require that this contribution is zero when the mass profile itself is zero at $r\to 0$. In other words, the field gradient should diverge slower than $r^2$ at the origin. In the case of a screened field, the integration constant is obtained by imposing continuity at the screening radius $\rc$: a match between the inner and the outer profiles and the first derivative(s).
%
\section{Solutions for different mass profiles}\label{sec:solutions}
%
   In this section, a brief description of the formalism to obtain the semi-analytic solutions of the field and its derivatives is provided, for the different mass density models considered here. For notation simplicity, $\mathcal{B}=\beta{\rhos \rs^2}/{\Mp}$ is introduced. Schematically, the procedure for computing the semi-analytic solution for any mass density model is as follows: i) obtain the exterior solution for the field as defined in \cref{eq:outer}, assuming spherical symmetry, which depends on two free parameters $\rc$ and $\Cs$, ii) these two free parameters are now obtained by fixing the boundary conditions at $r=\rc$ within which the field is assumed to be negligible. Now one can put together interior $\phi_{\rm int}(r)$ and the $\phi_{\rm ext}(r)$ to obtain the solution of $\phi(r)$ in the entire radial range of the object and outside. 
   
%
    \subsection{Solutions of the NFW-type}
%
    The expression for the chameleon field in the case of NFW and $b$NFW has already been presented in previous works (\textit{for instance,} \cite{Terukina12, Wilcox:2015kna, Pizzuti:2020tdl}). For the sake of brevity, we briefly review the spherical solution to the field outside the screening where $r>r_c$, further providing the matching conditions (\textit{i.e.} the screening equation) with the interior solution $\phi_\text{int} \simeq 0$. The field gradient, which enters in the expression of the fifth force is given by,
        \begin{equation}\label{eq:dphi_bnfw}
         \frac{\text{d}\phi_\text{ext}}{\text{d}r}=\frac{(x+1)^{1-b} (x-b x-1)}{x^2(b-1)(b-2)}\mathcal{B}+\frac{\Cs}{\rs x^2}\ . 
        \end{equation}
    Integrating the above expression yields the exterior solution to the field profile as, 
        \begin{equation}
         \phi_\text{ext}(x)=\frac{(1+x)^{2-b}}{x(b-1)(b-2)}\mathcal{B}-\frac{C_\text{s}}{x}+\phi_\infty\ .
        \end{equation}
    {Here $\phiinf$ denotes the asymptotic value of the chameleon field.} Assuming that the field is negligible ($\phi_\text{int}\simeq 0$) within the interior (\textit{i.e.} $r< \rc$), and enforcing the continuity between the exterior and interior solutions at $\xc \equiv \rc /\rs$ one obtains, 
        \begin{equation}\label{eq:sqbNFW} 
            \begin{split}  
             & \xc = \left(\frac{\phi_\infty(b-1)}{\mathcal{B}}\right)^\frac{1}{1-b}-1\ , \\
             & \Cs=\frac{(1 + \xc)^{1-b}(1-\xc+b\,\xc)}{(b-1)(b-2)}\mathcal{B}\ . \\
            \end{split}
        \end{equation}
    Amongst the above equations, the solution to the former, which we term as screening function ($\rm{f_s(r)}$), provides us with the screening radius ($\rc$). In the case where the field is not screened, the exterior solution holds at any $r>0$, with the integration constant given by,
        \begin{equation}
         C_\text{s}=\frac{\mathcal{B}}{(b-1)(b-2)}\ ,
        \end{equation}
    which is strictly valid for $b>2$. For the $b=2$ case, the NFW case, the exterior field gradient:
        \begin{equation}
         \dr{\phi_\text{ext}}=\frac{\mathcal{B}}{r_\text{s}\, x^2}\left[\frac{1}{x+1}+\ln (x+1)\right]+\frac{C_\text{s}}{r_\text{s}\, x^2}\ ,
        \end{equation}
    and the field profile:
        \begin{equation}
         \phi_\text{ext}(x)=-\, \frac{1+\ln(x+1)}{x}\mathcal{B}-\frac{C_\text{s}}{x}+\phi_\infty\ ,
        \end{equation}
    with the junction conditions at the matching radius, $\xc$\footnote{Note that the set of equations differs from the one of \cite{Terukina:2013eqa} due to a different definition of the normalization constant $C_\text{s}$.} 
        \begin{equation} \label{eq:NFWscreen}
         \xc=\left[\frac{\mathcal{B}}{\phi_\infty}-1\right]\,,  \quad \Cs=-\phi_\infty-\mathcal{B}\ln\left(\frac{\mathcal{B}}{\phi_\infty}\right)\ .
        \end{equation}
    The unscreened solution is obtained by imposing $\Cs=-\mathcal{B}$. Finally, the field profile associated with the $g$NFW matter density comes as:
        \begin{equation}
            \begin{split}
             \dr{\phi_\text{ext}}= \frac{\mathcal{B}}{\rs x^2}\bigg[\frac{x^{3-\gamma}}{3-\gamma}\,{_2}F_1(3-\gamma,3-\gamma,4-\gamma,-x)
             + 1\bigg]+\frac{\Cs}{\rs^2x^2}\ ,
            \end{split}
        \end{equation}
    and 
        \begin{equation}
            \begin{split}
             \phi_\text{ext}(x) = -\frac{\mathcal{B}}{x}\left[ \frac{x^{3-\gamma}(1+x)^{\gamma-2}-2-x+\gamma}{\gamma-2}
              + \frac{x^{3-\gamma}}{3-\gamma}\,{_2}F_1(3-\gamma,3-\gamma,4-\gamma,-x)  \right]
              -\frac{C_\text{s}}{r_\text{s}\, x}+\phi_\infty\ .
            \end{split}
        \end{equation}
    with the junction conditions given by
        \begin{equation}
            \begin{split}
             x_c & = \left[1-\left( 1+ \left( \frac{\phi_\infty}{\mathcal{B}}\right)(\gamma -2)\right)^{1/(2-\gamma)}\right]^{-1} -1 \,,  \label{eq:sqNFW}  \\
             \Cs &=- \mathcal{B} \rs \left[ \frac{\xc^{3-\gamma}}{3-\gamma}\,{_2}F_1(3-\gamma,3-\gamma,4-\gamma,-\xc) + 1 \right] \,. \\    
            \end{split}
        \end{equation}
    When the field is in the unscreened regime, no real positive solutions for the screening radius $\rc$ can be found. The integration constant of the exterior field \cref{eq:outeradial} is then given by $\Cs= - \mathcal{B}\, \rs$. It is worth pointing out that in \cref{eq:sqNFW,eq:sqbNFW}, $\rc$ can be explicitly expressed as a function of $\phi_\infty$. Such a relation is, however, not straightforward for all models. 

%
    \subsection{Burkert solutions}\label{sec:Bur}
\begin{figure}
         \includegraphics[scale=0.45]{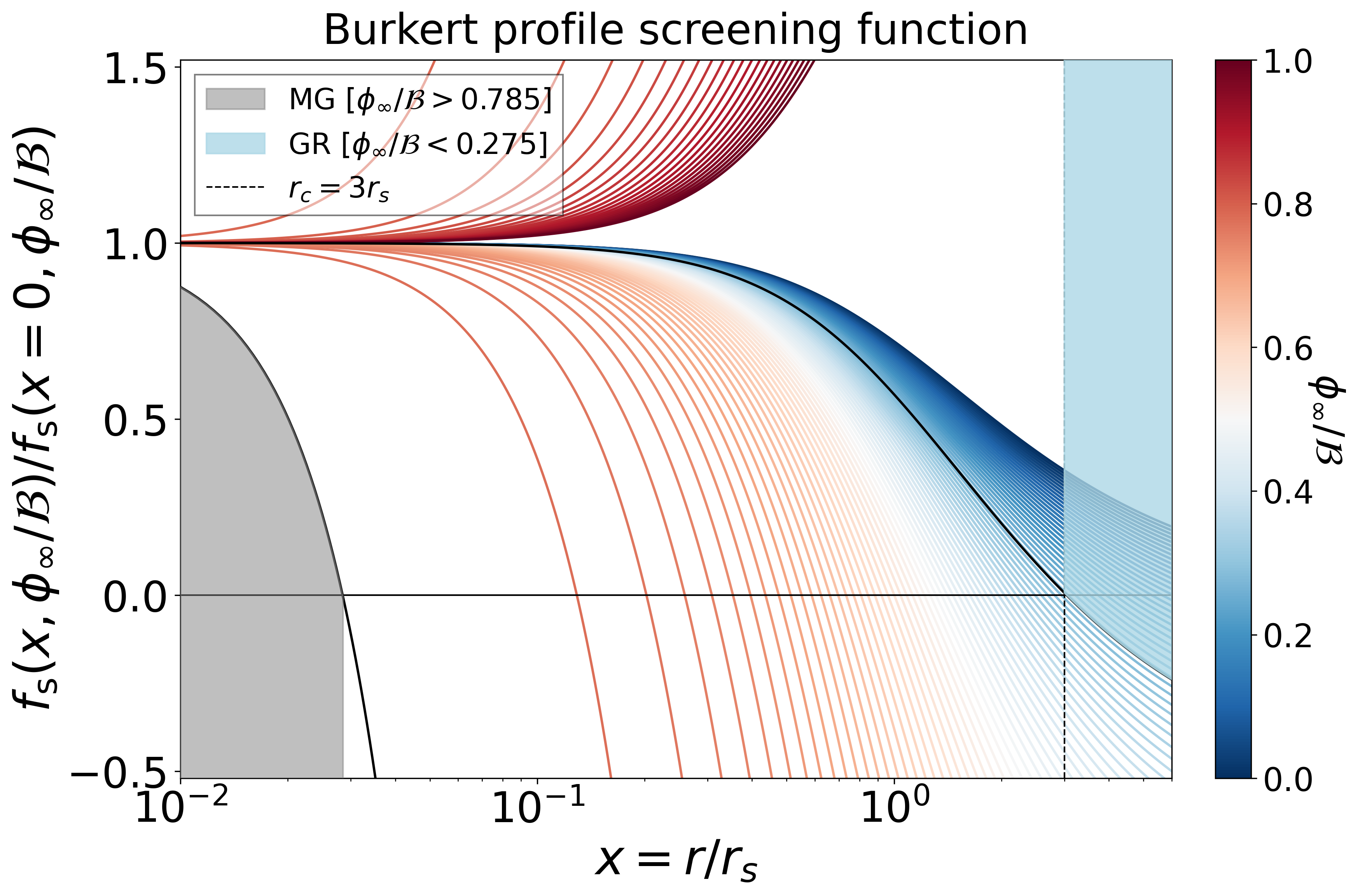}
         \includegraphics[scale=0.45]{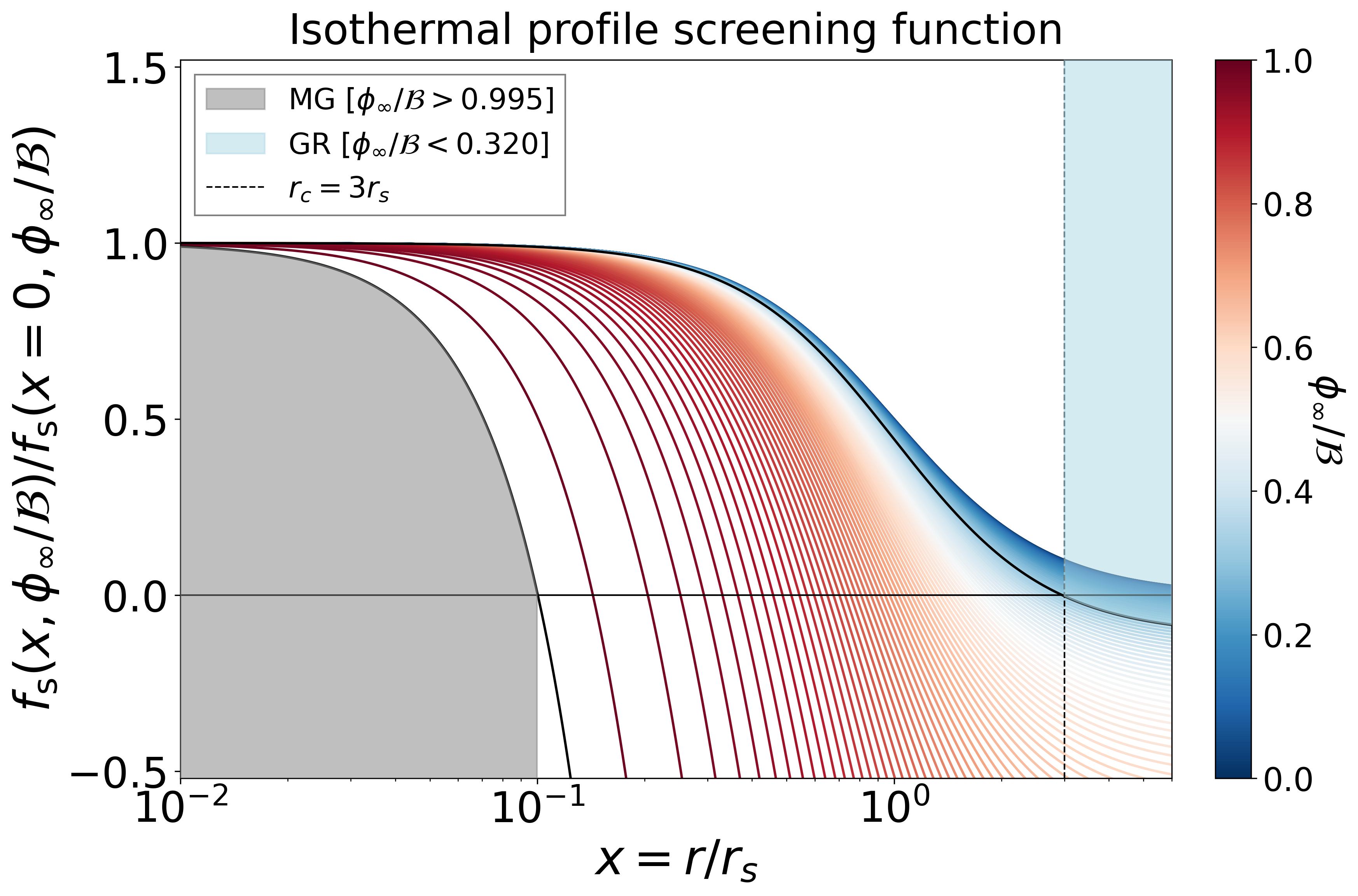}
         \caption{Variation of the screening function ($\rm{f_s (x)}$) as a function of $\frac{\phi_\infty}{\mathcal{B}}$, for Burkert (\textit{top}) and Isothermal (\textit{bottom}) profiles. The vertical dashed line marks the solution when the screening radius $\rc = 3 \rs$ for a given value of $\phiB$ for the respective mass models. The intersection of each profile with the $\rm{f_s (x)} =0$, provides the value of $\rc$, within which the effects of the Chameleon field are screened. To avoid the sign ambiguity of the screening function we show the re-scaled $\text{f}_\text{s}(r)/\text{f}_\text{s}(r = 0)$.}
         \label{fig:scrf}
    \end{figure}

%
    For the Burkert model, the exterior field gradient is:
        \begin{equation}
            \begin{split}
             \dr{\phi_\text{ext}}= \frac{C_\text{s}}{r_\text{s}\, x^2}+\frac{\mathcal{B}}{r_\text{s}\, x^2}\Big[\frac{1}{4} \ln \left(x^2+1\right) 
              +\frac{1}{2} \ln (x+1)-\frac{1}{2} \tan ^{-1}(x)\Big]\ ,
            \end{split}
        \label{eq:burket grad}
        \end{equation}
    and the field profile can be written as,
        \begin{equation}
            \begin{split}
             \phi_{\text{ext}} (x)= & -\frac{C}{x} + \frac{\mathcal{B}} {4\, x} \bigg[(x-1) \ln\big(x^2+1\big) \\ & +2 (x+1) \Big(\tan ^{-1}(x)-\ln (x+1)\Big)\bigg]\\
             &-\frac{\pi }{4}\mathcal{B} +\phi_\infty\ .
            \end{split}
            \label{eq:burket s}
        \end{equation}
    Note that the factor $\mathcal{B}\, \pi/4$ ensures $\phi \to \phi_\infty$ for $x\to \infty$, and the matching with the inner solution is obtained when
        \begin{equation*}
         C_\text{s}= \frac{1}{4} \mathcal{B} \left[-\log \left(\xc^2+1\right)-2 \log (\xc+1)+2 \tan ^{-1}(\xc)\right]         
        \end{equation*}
    Finally, the screening equation is given as, 
        \begin{equation}
         \ln \left[\frac{\xc^2+1}{(\xc+1)^2}\right]+2 \tan ^{-1}(\xc)=\pi -4 \frac{\phi_\infty}{\mathcal{B}}\,, \label{eq:sa Burket}
        \end{equation}
    where the matching conditions to get $r_c$ cannot be solved analytically. In the top panel of Fig.~\ref{fig:scrf} we show the screening function $\rm{f_s} (\xc)/{\rm f_s} (0)$, with
    \begin{equation*}
        {\rm f_s} (\xc) = \ln \left[\frac{\xc^2+1}{(\xc+1)^2}\right]+2 \tan ^{-1}(\xc) - \pi + 4 \frac{\phi_\infty}{\mathcal{B}} \,,
    \end{equation*}
    plotted for varying values of $\phiB$. Wherein each profile has been normalised to its value at $r =0$. The intersection of the screening function profiles at $\rm{f_s (\rc)} = 0$ provides the screening radius $\rc$ for a given value of $\phiB$. It is illustrative to notice that for $\phiB \gtrsim 0.785 \sim \pi/4$, there exists no solution to the screening function and the entire cluster is unscreened. As it can also be seen by the structure of \cref{eq:sa Burket}, the solution for the screening radius $r_c$ demands $\frac{\phi_\infty}{\mathcal{B}} \leqslant \pi/4$, \textit{i.e.} the \textit{rhs} should be greater or equal to zero since the \textit{lhs} is a monotonically-increasing function of $x_c =r_c/r_s$, which is zero for $r_c=0$, thus, strictly positive for $x_c>0$. As mentioned earlier, this corresponds to the case where the field is unscreened and the entire cluster experiences the fifth force.

    \subsection{Isothermal solutions}
%
   The Isothermal mass density model's field gradient is
        \begin{equation}            \label{eq:Iso grad}
             \dr{\phi_\text{ext}}= \left[C_\text{s}-\mathcal{B}\left(\frac{x}{\sqrt{x^2+1}} - \ln \left(\sqrt{x^2+1}-x\right)\right)\right]\frac{1}{r_\text{s} x^2}\ ,
        \end{equation}
    which, after integration, results in the following field profile
        \begin{equation}\label{eq:iso phi}
         \phi_{\text{ext}}(x)=- \frac{\ln(\sqrt{x^{2}+1}+x)}{x}\mathcal{B}-\frac{C_\text{s}}{x}+\phi_{\infty}\ ,
        \end{equation}
    where $C_\text{s}=\phi_{\infty}\, \xc-\mathcal{B}\ln(\sqrt{\xc^{2}+1}+x_{c})$. Repeating the process explained in \cref{sec:solutions}, the screening equation is given by,
        \begin{equation}
         \sqrt{\xc^{2}+1}=\frac{\mathcal{B}}{\phi_{\infty}}\ ,
         \label{screening eq isothermal}
        \end{equation}
     which has a solution only when $ (\mathcal{B}/\phi_{\infty})^{2}>1$.The resulting screening radius is
        \begin{equation}\label{rc isothermal}
         \rc=\rs \sqrt{\left(\frac{\mathcal{B}}{\phi_{\infty}}\right)^{2}-1}\ .
        \end{equation}
    Note that in the unscreened regime (\textit{i.e.} $\xc = 0$) the constant vanishes, $C_\text{s} = 0$. In the bottom panel of Fig.~\ref{fig:scrf} we show the screening function profiles for the Isothermal mass density profile corresponding to \cref{rc isothermal}, $\text{f}_\text{s}(\xc) = \sqrt{\xc^{2}+1} -\mathcal{B}/\phi_{\infty}$. In contrast to the Burkert profile, we find that there always exists a solution to the screening function even in the limit $\phiB \to 1$. 
%
    \subsection{Einasto solutions}
%
    The gradient of the Chameleon field for the Einasto mass density model is given by
        \begin{equation}
         \dr{\phi_\text{ext}}=\frac{C_\text{s}}{r_\text{s}\, x^2}-\mathcal{B}\, 8^{-m} e^{2 m} m^{1-3 m} \frac{\Gamma \left(3 m,2 m\, x^\frac{1}{m}\right)}{r_\text{s}\, x^2}\ ,
        \end{equation}
    and
        \begin{equation}
            \begin{split}
             \phi_\text{ext}(x)= & \phi_\infty-\frac{C_\text{s}}{x}+\frac{\mathcal{B}}{x} 8^{-m} e^{2 m} m^{1-3 m} \times \\
             & \left[2^m m^m x\, \Gamma \left(2 m,2 m\, x^\frac{1}{m} \right)-\Gamma \left(3 m,2 m\, x^\frac{1}{m}\right)\right]\ .
            \end{split}
        \end{equation}
    Where $\Gamma(n,z)$ is the upper incomplete gamma function:
        \begin{displaymath}
         \Gamma(n,z)=\int _z^{\infty }\text{d} t \, t^{n-1} e^{-t}\ .
        \end{displaymath}
    Despite the complicated look of the field profile, it is still possible to obtain an analytical solution for the junction conditions between the screened and unscreened regimes:
        \begin{equation}
            \begin{split}
             & \rc=\rs \left[\frac{1}{2 m}Q^{-1}\left(2 m,\frac{4^m \phi_\infty m^{2 m-1}}{\mathcal{B}\, e^{2 m}\, \Gamma (2 m)}\right)\right]^m\ ,  \\
             & \Cs=\, 8^{-m} e^{2 m} m^{1-3 m} \Gamma \left(3 m,2 m\, x_\text{s}^{\frac{1}{m}}\right)\mathcal{B}\ , \\
            \end{split}
        \end{equation}
    where $Q^{-1}(y,a)$ is the inverse of the upper regularized incomplete gamma function, as well as for the case where there is no screening
        \begin{equation}
         \Cs= 8^{-m} e^{2 m} n^{1-3 m}\, \Gamma (3 m)\mathcal{B}\ .
        \end{equation}
%
    \subsection{Solution's existence}
%
    Finally, let us discuss the solutions' existence and compatibility with the Chameleon screening mechanism through their asymptotic behaviour. It is clear from the previous mass models \cref{eq:NFW,eq:bNFW,eq:gNFW,eq:bur,eq:Iso,eq:Ein} that when $r/\rs \gg 1$, \cref{eq:outer} becomes 
        \begin{equation}
         \frac{1}{r^{2}}\frac{d}{dr}\left(r^{2}\frac{d}{dr}\phi\right)\sim\frac{1}{r^{3}}\ ,
        \end{equation}
    with asymptotic solution 
        \begin{equation}
         \phi\sim\frac{C}{r}+\phi_{\infty}\ .
        \end{equation}
    Therefore, all considered mass models are compatible with the Chameleon screening mechanism; that is, the field converges asymptotically to a finite background value $\phi_\infty$. Note that this wouldn't be the case if the matter density at large $r$ goes as $1/r^2$. In that case, the field's solution does not converge to a finite value. In order to demonstrate that, assuming that the mass model goes as $1/r^\alpha$ for large $r$ one can show that the resulting asymptotic behaviour is
        \begin{equation}\label{eq: asympt alpha}
         \phi\sim\frac{C}{r}+C'+\frac{1}{(1-\alpha)(2-\alpha)}r^{2-\alpha}\ ,
        \end{equation}
    which consequently requires $\alpha >2$ to have a finite solution.

        \begin{figure}
    \centering
         \includegraphics[scale=0.5]{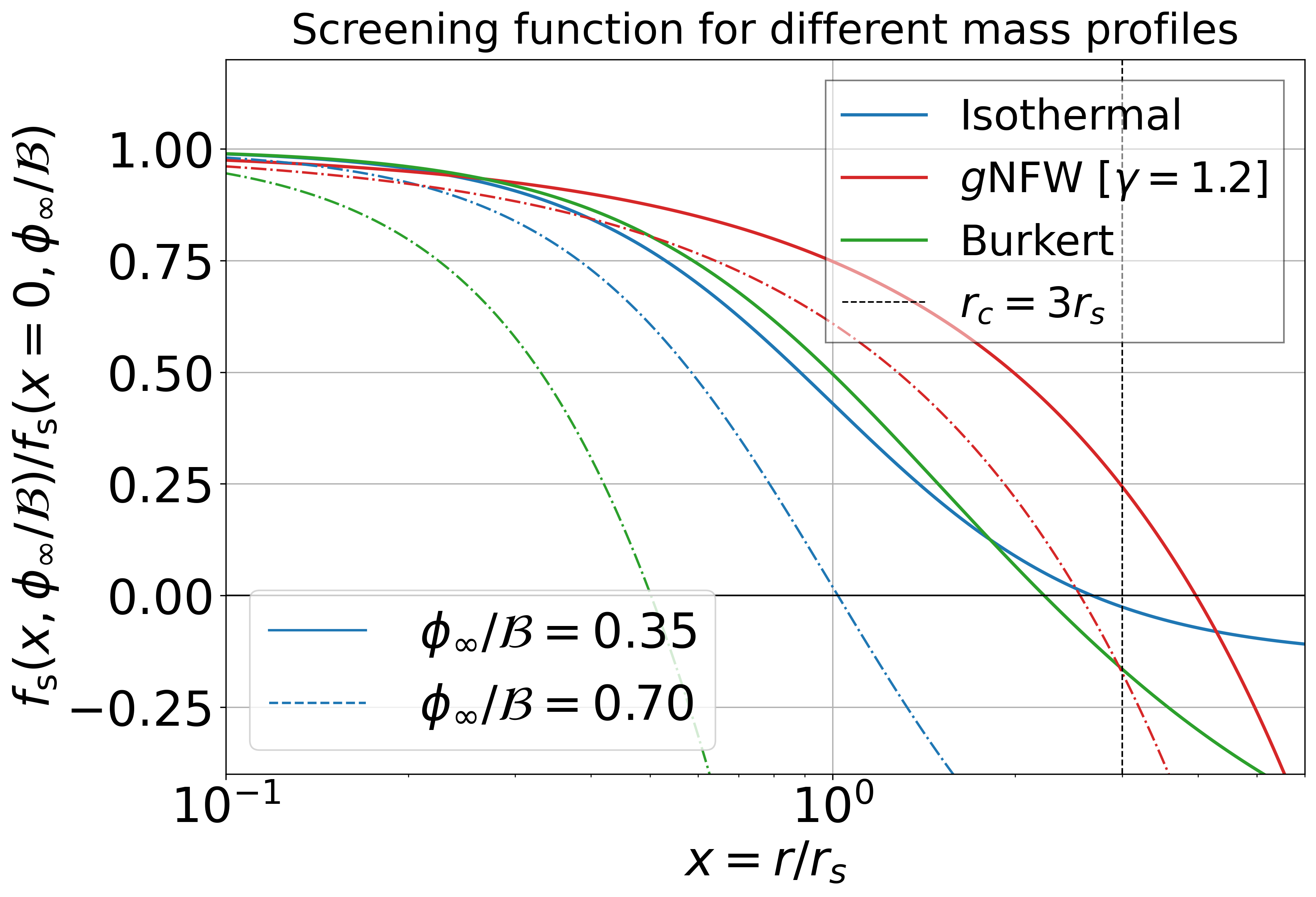}
         \caption{Comparison of the screening function for three different mass models, namely, Isothermal, gNFW and Burkert are shown. The screening functions plotted for two different values of $\phiB = 0.35$ (solid) and $\phiB = 0.7$ (dashed) are compared.}
         \label{fig:scrf_comp}
    \end{figure}

    Finally, in Fig~\ref{fig:scrf_comp}, we show the comparison of the screening functions' behaviour for different mass models. Let us compare amongst the mass models shown for the same value of $\phiB$; we can immediately notice that the $g$NFW profile provides larger $\rc$ with respect to the Isothermal and Burkert mass profiles. This implies that for the same normalization set by $\phiB$, using the $g$NFW mass profile more of the cluster is screened. It is also interesting to note that the dependence of the solution to the screening varies significantly in the Isothermal and Burkert cases in comparison to the $g$NFW case. 

    This, in turn, implies that the $g$NFW profile coupled with the chameleon field screens the fifth force effects within the galaxy cluster more effectively than the two other mass profiles in comparison here. For the purpose of illustration, here we show the $g$NFW profile assuming $\gamma =1.2$. Note that $r_c$ grows with $\gamma$, and it remains larger than that obtained in the case of Burkert or Isothermal profiles for the same value of $\phi_\infty/\mathcal{B}$ down to the lower limit of $\gamma = 0$. This further indicates that the difference in $r_c$ will be even larger for $\gamma \to 2$ in comparison to $\gamma = 1.2$.
%
 \section{Comparison with numerical solutions}\label{sec:numerical}
%
    In order to validate the approach described in \cref{sec:solutions}, let us compare the obtained semi-analytical solutions with the numerical solution of Eq.~\ref{eq:eq_of_motion_s}.

    The set of numerical solutions of Eq.~\ref{eq:eq_of_motion_s} were obtained through a $6^{th}$-order explicit Runge-Kutta integrator while the appropriate boundary conditions -- $\nabla ^2 \phi \approx 0 $ at the centre of the mass distribution, and $\text{d}\phi /\text{d}r =0$ at infinity -- were imposed through a Newton-Rapshon shooting method. In order to avoid the divergence at the centre of the mass density distribution associated with some of the models, an inner cutoff radius was imposed. The value of the latter ranged between $5-10\%$ of the screening radius in order to get the best fit for the semi-analytic approach. The appropriate boundary condition at infinity was imposed by considering a numerically small value of the scalar field derivative, $\sim 10^{-8}$, at a scaled radius $x$ several times larger than the main mass distribution, $x_{max}\approx 10^3$. 

\begin{figure}
     \centering
     \includegraphics[width =1.\columnwidth]{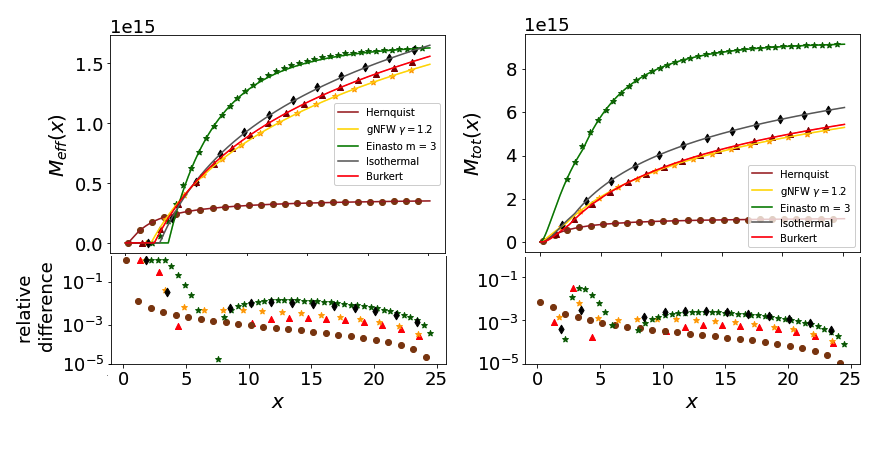}
     \vspace{-0.6in}
     \caption{Left: Semi-analytic approximation for the effective mass $M_\text{eff}$ (solid lines) as a function of $x=r/\rs$ compared with the numerical solution (points) for the different mass ansatz defined in Section \ref{sec:mass_models}. The bottom plot indicates the relative difference between the two. Right: the same for the total dynamical mass $M_\text{tot} = M+ M_\text{eff}$. The adopted parameters are $\rho_{s}= 5\times 10^{14}\,\text{M}_\odot/\text{Mpc}^3$, $r_{s}= 0.5\,\text{Mpc}$, $\beta = 0.5$, $\phi_\infty = 5\times 10^{-5}$.}    
     \label{fig:numerical}
    \end{figure}

    Comparative results between the semi-analytic (solid) and the full numerical solutions (points) can be seen in Fig.~\ref{fig:numerical} for all the mass density models under consideration, assuming $\rho_{s}= 5\times 10^{14}\,\text{M}_\odot/\text{Mpc}^3$, $r_{s}= 0.5\,\text{Mpc}\,$, $\beta = 0.5$ and  $\phi_\infty = 5\times 10^{-5}$. 

    Let us now quickly analyse the difference between the semi-analytical solution and the full non-linear numerically obtained solution, Fig.~\ref{fig:numerical}. Both the effective mass (top left) and the total mass (top right), as well as the relative difference between the semi-analytically obtained and the numerically obtained solutions (bottom), are represented in Fig.~\ref{fig:numerical}. For the study of how well the semi-analytical solution describes the true solutions, let us analyse the relative difference for the effective mass. As $x$ increases from the origin $x=0$ until the defined cut-off radius, $x=x_\text{s}$ -- where the strongest assumptions and approximations were made --, while the semi-analytical is set to zero, the numerical solutions are small but non-zero, originating a large relative difference which is accentuated at the transitional scaled radius $x_\text{s}$, where the numerical starts to gain significant non-zero values before the semi-analytical (smoother transition from a negligible value). This behaviour is, however, cancelled as one goes away from the mass distribution (increase $x$) to the background configuration (no mass distribution, flat scalar field profile). At this point, the semi-analytical and the numerically obtained solutions coincide almost perfectly, with a maximum relative error of $10^{-3}$. Hence, besides some slight differences in the screening radius transition and asymptotic behaviour, the semi-analytic approximation describes with a high degree of agreement the full numerically obtained results, giving confidence for their use in more complex calculations that will proceed. Note that the maximum discrepancy between the numerical and semi-analytic approach in the total mass $M(r)+ M_\text{eff}(r)$ (right plots) is of $7\%$, and corresponds to $x = \sim 2$ (\textit{i.e.} $r \sim 1\, \text{Mpc}$ for the adopted value of $\rs$).
%
\section{Kinematics mass reconstruction: simulations and testing}\label{sec:kinematics}
%
    Having established the algebraic expression for the fifth force due to the Chameleon coupling with several mass densities, some of the possible systematic effects induced by the choice of a `wrong' mass model when reconstructing the mass profile of clusters in modified gravity are explored. For this purpose, we focus on mass determination using kinematic analyses of member galaxies, assuming dynamical relaxation and spherical symmetry, with additional priors on the mass profile parameters simulating the constraint obtained from a lensing-like reconstruction of the cluster phase space.

    Using an upgraded version of the \textsc{ClusterGEN} code (see \cite{Pizzuti:2020tdl}), we generate synthetic, spherically symmetric, isolated systems of collisionless particles in dynamical equilibrium. The particles are distributed according to an NFW number density profile $\nu(r)=\nu_\text{NFW}(r,r_\nu)$, with a characteristic scale radius $r_\nu = 0.5\, \Mpc$\footnote{The distribution of the member galaxies in clusters can be different from that of the total matter density, which accounts for several mass components (see \textit{e.g.} \cite{Budzynski2012, Mamon2019}). Here, we consider the NFW ansatz for simplicity, but no significant changes have been found when adopting other density models.}. The velocity field follows a dispersion along the radial direction, $\sigma^2_r$, obtained by solving the \textit{Jeans' equation}:
    \begin{equation}\label{eq:jeans}
     \frac{\text{d} (\nu \sigma_r^2)}{\text{d} r}+2\eta(r)\frac{\nu\sigma^2_r}{r}=-\nu(r)\frac{\text{d} \Phi}{\text{d} r}\ ,
    \end{equation}
    where $r$ is the 3-Dimensional radial distance from the cluster centre, $\nu(r)$ corresponds to the number density profile mentioned above, $\Phi$ is the total gravitational potential and $\eta \equiv 1-(\sigma_{\theta}^2+\sigma^2_{\varphi})/2\sigma^2_r$ is called velocity anisotropy, where $\sigma_{\theta}^2$ and $\sigma^2_{\varphi}$ are the velocity dispersion components along the tangential and azimuthal directions, respectively. In spherical symmetry we have $\sigma_{\theta}^2=\sigma^2_{\varphi}$ and the expression of the anisotropy profile simplifies to $\eta = 1 -\sigma_{\theta}^2/\sigma_{r}^2$. In our simulation $\eta(r)$ is modeled by a Tiret profile \cite{Tiret01}:
    \begin{equation} \label{eq:anisT}
     \eta(r) = \frac{r}{r+r_{\beta}}\eta_\infty\ ,
    \end{equation}
    where $r_{\beta}$ is a scale radius equivalent to $r_{-2}$ of the assumed mass model, and the value of the anisotropy at infinity, $\eta_\infty$, is set to $0.5$. \cref{eq:anisT} is found to provide an adequate fit for the velocity anisotropy of cluster-size halos in cosmological simulations (\textit{e.g.} \cite{Mamon_2005}).

\begin{figure*}
    \centering
     \includegraphics[scale=0.3]{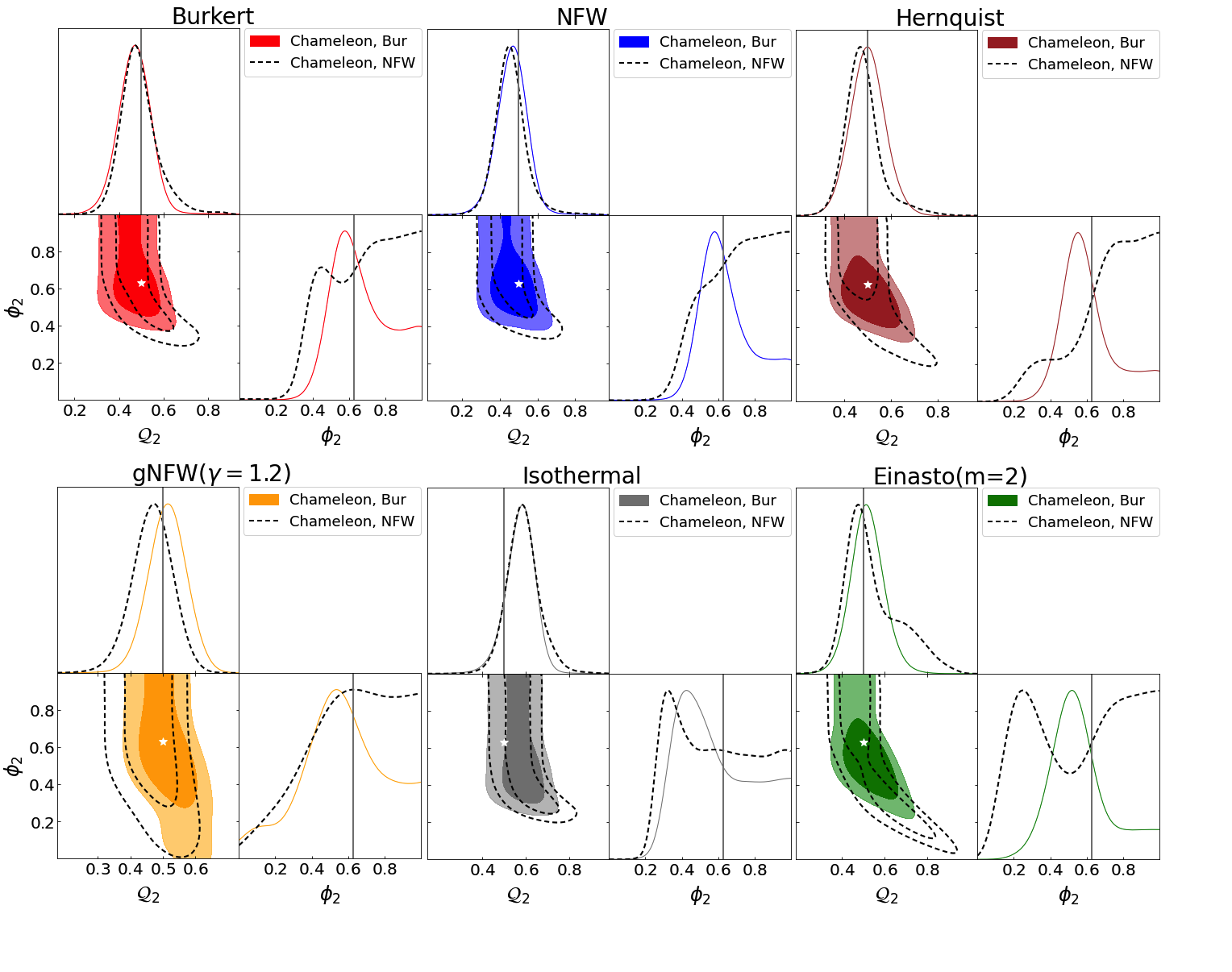}  
     \vspace{-0.6in}
     \caption{Marginalized posteriors for the one and two-dimensional distributions of $\phi_2$ and $\mathcal{Q}_2$  from the \textsc{MG-MAMPOSSt} analysis of two clusters generated in Chameleon gravity (case II) assuming a Burkert profile (solid shaded areas) and a NFW profile (dashed contours). Each plot refers to a different mass density ansatz used in the fit. Inner and outer shaded regions, as well as inner and other dashed contours, refer to 1-$\sigma$ and 2-$\sigma$ limits, respectively. The corresponding parameter constraints are presented in \cref{tab:resMG}. The simulated cluster is represented by $*$ in each panel, corresponding to $\{\phi_2, \mathcal{Q}_2\} = \{0.63, 0.5\}$.}
     \label{fig:case2}
    \end{figure*}

    Each cluster is populated up to 7 times the virial radius $r_{200}$, considering 300 particles within $r_{200}$; this corresponds to a realistic number of spectroscopic redshift that current and upcoming surveys \cite{Euclid:2019bue} can provide for a few dozens of clusters. Three different parametrizations of the gravitational potential are adopted: two in Newtonian gravity, namely a standard NFW and a Burkert model, and two in Chameleon gravity, namely an NFW profile and a Burkert model, assuming for both cases $\phi_\infty/\Mp=  10^{-4}\,c^2$ and $\beta=1.0$\footnote{These values for the physical parameters correspond to $\phi_2 \equiv 1-\exp(-\phi_{\infty}/\Mp [10^{-4} c^2]^{-1}) \simeq 0.63$ and $\mathcal{Q}_2 \equiv \beta/(1+\beta) = 0.5$, for which the results are presented in Fig.~\ref{tab:resMG} and Fig.~\ref{fig:case2}.}, corresponding to very strong departures from GR. This is to analyze the effect of the screening mechanism in halos of the same size but with different shapes living in a $\Lambda$CDM e in a modified gravity scenarios, respectively. For all cases, we consider $r_{200}= 1.6\,\text{Mpc}$, $r_{-2} = 0.5 \,\text{Mpc}$. The latter is defined as the radius at which the logarithmic derivative of the density profile is equal to $-2$,
    \begin{equation}
     \left(\frac{\text{d}\ln\rho(r)}{\text{d}\ln r}\right)_{r_{-2}} =-2\ ,
    \end{equation}
    which can be connected to the scale radius, $r_{-2}$, of each density profile in \cref{eq:NFW,eq:gNFW,eq:bNFW,eq:bur,eq:Iso,eq:Ein}. In particular, $r_{-2}=r_\text{s}$ for the NFW and Einasto models, while $r_{-2}\simeq 1.52\,r_\text{s}$ for Burkert, $r_{-2} = \sqrt{3}\,r_\text{s}$ for Isothermal, $r_{-2} = r_\text{s}/(b-1)$, for b-NFW and $r_{-2} =(2-\gamma)r_\text{s}$ for $g$NFW. 

    The total gravitational potential of our synthetic clusters is then reconstructed by means of \textsc{MG-MAMPOSSt}\footnote{Code is publicly available at \href{https://github.com/Pizzuti92/MG-MAMPOSSt}{https://github.com/Pizzuti92/MG-MAMPOSSt}.} program package (see \cite{Pizzuti:2020tdl, Pizzuti:2022ynt}), a modified version of the \textsc{MAMPOSSt} code \cite{Mamon01}, which determines the mass, anisotropy and number density profiles of spherically symmetric systems by solving the Jeans' analyses. Given parametric input models of anisotropy profile ($\eta(r)$), number density ($\nu(r)$),  and the gravitational potential ($\Phi(r)$), the code performs a Monte Carlo Markov-Chain (MCMC) sampling of the parameter space, using as input data the line-of-sight velocity field, $v_z$, and the projected positions, $R$, of the member galaxies. The dataset $(R,v_z)$ is called projected phase space (pps), which is our final simulated observable. In particular, \textsc{MG-MAMPOSSt} computes the probability $q(R_i,v_{z,i}| \theta)$ of finding a galaxy at the point $R_i,v_{z,i}$ in the pps. The (log) likelihood is then given by $\mathcal{L}_\text{dyn}(\theta) = \sum_i \text{ln} q(R_i,v_{z,i}| \theta)$.

\begin{figure}
        \centering
        \includegraphics[width=0.9\linewidth]{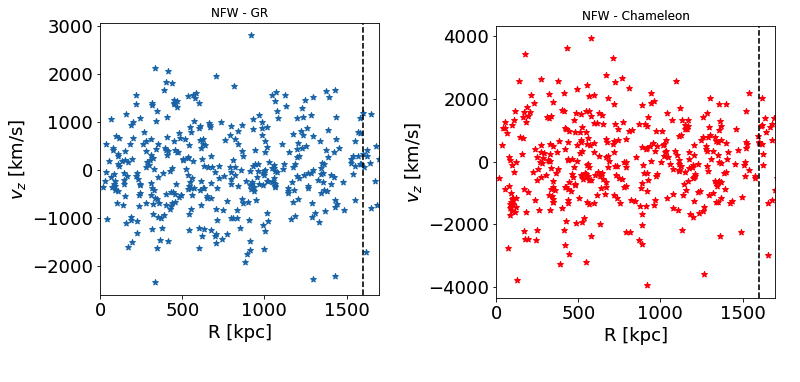}  
        \vspace{-0.2in}
        \caption{Projected phase spaces for the NFW-GR cluster (left) and the NFW cluster in Chameleon gravity (right). The black dashed line indicates the value $R = r_{200} \equiv 1.6 \, \text{Mpc}$.}
        \label{fig:pps}
    \end{figure}

    Figure \ref{fig:pps} shows the pps for the cluster generated with the NFW profile in GR (left) and in Chameleon gravity (right). As one can already visually inspect, the velocity dispersion of the cluster in modified gravity scenario tends to be larger than the one in GR, due to the additional contribution of the fifth force.
    
    \textsc{MG-MAMPOSSt} requires also information about the number density profile $\nu(r)$ of the galaxies, which can be generally obtained externally from \textsc{MG-MAMPOSSt} by fitting the observed projected distribution of galaxies in clusters (see \textit{e.g.} \cite{Biviano:2023oyf, Sartoris2020}). Thus, the scale radius $r_\nu$  of the number density profile can be considered as a fixed parameter in the MCMC, with the value provided by the external fit;  note that this is independent of the model adopted for the total mass profile in the Jeans' equation. Here by construction, the mock pps are built in such a way fitting the number density of galaxies with a NFW model will provide the correct value of the scale radius $r_\nu = 0.5\, \text{Mpc}$. It is important to mention that variation of the constraints on the inferred dynamical mass in modified gravity, induced by variation of the number density scale radius $r_\nu$ are negligible (see \textit{e.g.} \cite{Pizzuti17}).
    
    As shown in \cite{Pizzuti:2020tdl}, kinematics data alone cannot constrain efficiently the parameter space of Chameleon models, due to the degeneracy between the parameters of the density profile $r_{200}$ and $r_\text{s}$ and the fifth force. As such, additional independent information should be provided on $r_{200}$ and $r_\text{s}$ in order to break the degeneracy. In this regard, cluster mass determination with gravitational lensing analyses is a powerful tool to be combined with kinematic data. Indeed, due to the conformal structure of Chameleon theories and conformal invariance of null geodesics, photons do not perceive the effect of the fifth force (see e.g. \cite{Burrage:2017shh}). We thus simulate the availability of a lensing-like information as a Gaussian distribution $P_\text{l} =\mathcal{G}(r_{200},r_\text{s})$. Despite all effects due to asphericity are neglected, as well as systematics in the lensing modeling (see \cite{Umetsu_2020} for a review), to determine the mean values $\bar{r}_{200},\bar{r}_\text{s}$ we should still account for the effect of choosing the wrong mass profile also in a lensing analysis. Indeed, even if the value of $r_{200}$ should be, in principle, the same when inferred by the wrong mass model (as this does not depend on the shape of the halo), the scale radius $r_\text{s}$ differs from profile to profile. Given the value of $r_{-2}$, which is fixed by the "true" mass profile, each mass model would infer a $r_\text{s}$ which is, in general, a function of $r_{-2}$ as discussed above. Thus, the central value of $r_\text{s}$ in the lensing Gaussian varies for each profile according to 
    $\bar{r}_\text{s}=f( r_{-2} = 0.5 \,\text{Mpc} )$.
     
    As for the standard deviations, those are based on reliable uncertainties given current mass lensing estimates (\cite{Umetsu:2015baa}): $\sigma_{r_{200}} = 0.1\,r_{200}$ and $\sigma_{\rs} = 0.3\,r_\text{s}$. One can also check that variations of the correlation do not produce significant changes in the results. With this into account, let us set the correlation as $\mu = 0.5$.
     
    The aim of this section is twofold: \textit{a)} determine whether a cluster, generated in $\Lambda$CDM, produces spurious detection of modified gravity when the mass is reconstructed in a Chameleon gravity framework assuming the wrong model for the matter density, hereafter referred to as case I. And \textit{b)} investigate how the shape of the total mass distribution influences the fifth force imprint of a cluster-size halo in a Chameleon universe in the joint kinematics+lensing analyses (\textit{aka} case II). Note that, in both cases, possible effects induced by the halo's triaxiality are ignored, and solely spherical symmetry is considered. Results are presented in terms of the re-scaled variables $\mathcal{Q}_2=\beta/(1+\beta)$ and $\phi_2 = 1- \exp\left[\phi/(10^4\,\Mp)\right]$, which spawn the range $[0,1]$. Within this range, we consider uninformative uniform prior for both parameters, as already done in previous works (see e.g. \cite{Wilcox:2015kna,Boumechta:2023qhd}). As for the anisotropy parameter $\eta_\infty$, as a standard procedure (\textit{e.g.} \cite{Biviano:2023oyf}) we consider flat priors in the range $[0.5,7]$ in the parameter $\mathcal{A}_\infty = (1-\eta_\infty)^{-1/2}$.
    
    A sampling of 110000 points of the joint log-likelihood
    \begin{equation}
        \mathcal{L}_\text{tot} =  \mathcal{L}_\text{dyn} + \ln [ P(r_{200},r_\text{s})]\,,
    \end{equation}
    in the parameter space, assuming the above mentioned priors, is performed through a Metropolis-Hastings algorithm. We discard the first 10000 points as a burn-in phase, resulting in a final chain of 100000 samples. The convergence is ensured for each run by performing $n = 5$ chains and computing the corresponding Gelman-Rubin diagnostic coefficients $\hat{R}$ \cite{Gelman92}, checking that the requirement $\hat{R} \lesssim 1.1$ is always satisfied.

\begin{figure*}
    \centering
     \includegraphics[scale=0.3]{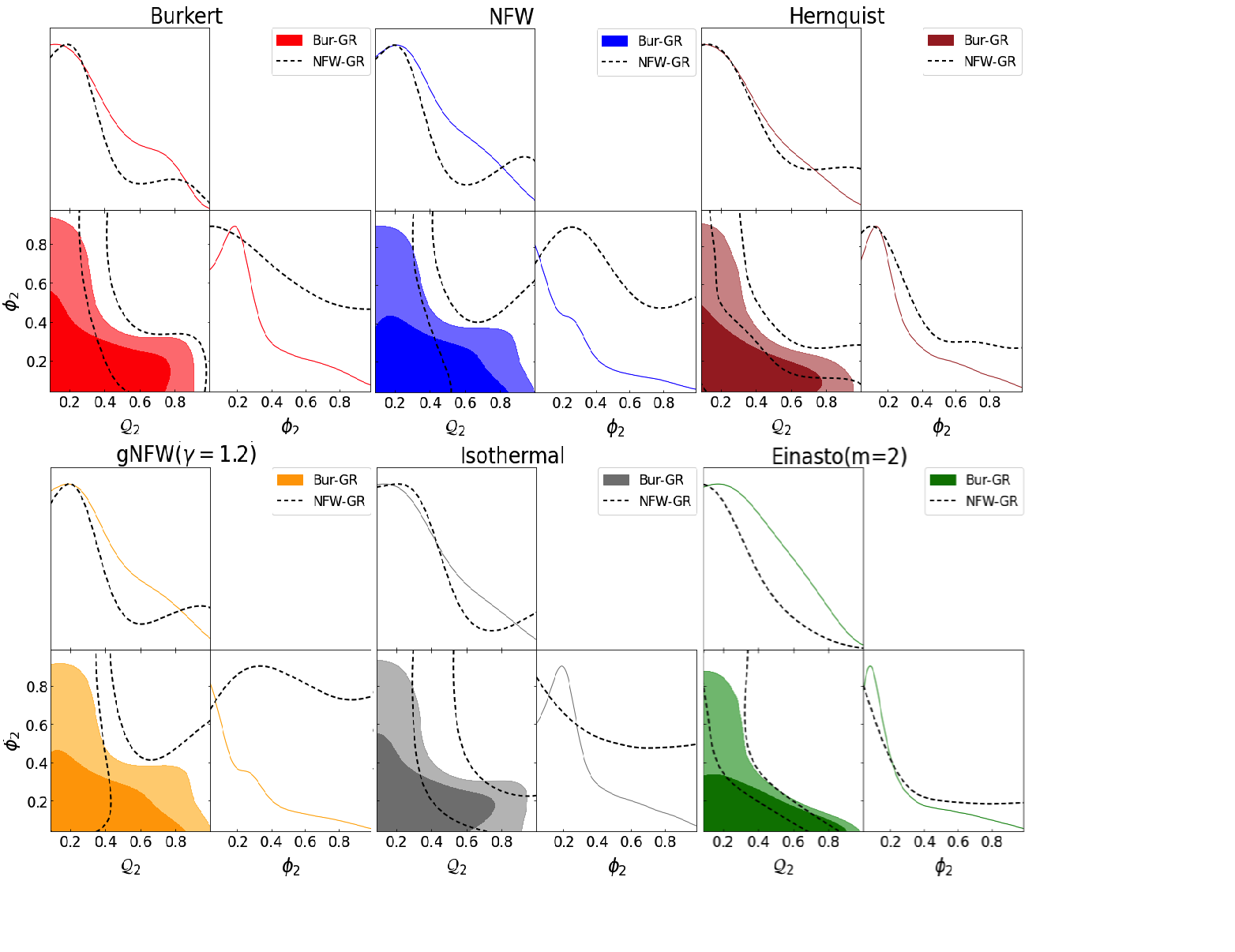}  
     \vspace{-0.6in}
     \caption{Marginalized one and two-dimensional distributions of $\phi_2$ and $\mathcal{Q}_2$ reconstructed by applying the \textsc{MG-MAMPOSSt} method to the GR clusters (case I). Shaded regions refer to a cluster generated with a Burkert profile, while the dashed contours are for a cluster generated assuming a NFW model.  Each plot refers to a different mass density ansatz used in the \textsc{MG-MAMPOSSt} fit. Inner and outer shaded regions (or inner and outer dashed contour lines) in the contour plots refer to one and 1-$\sigma$ contours. The plots are made by using the \href{https://getdist.readthedocs.io}{getdist package}, see also \cite{lewis2019}.}
     \label{fig:case1}
    \end{figure*}

    Fig.~\ref{fig:case1} shows the marginalized distributions of $\mathcal{Q}_2$ and $\phi_2$ in case I. The darker and lighter shaded areas indicate the one and two-$\sigma$ regions allowed in the parameter space for the Burkert-generated halo while the inner and outer dashed contours are the same for the NFW-generated halo. No significant shifts towards $\phi_2\ne 0$ and $\mathcal{Q}_2\ne 0$ are observed when a ``wrong'' mass model is used to estimate the gravitational potential independently of the mass density model. This indicates a sub-dominance of the possible biases due to the mass modelling when compared to the statistical uncertainties, suggesting the robustness of our method against these specific systematics. Nevertheless, the relation between the efficiency of the screening mechanism and the chosen density profile is clearly visible in the shape of the distributions. Models like Burkert, Isothermal or Einasto tend to have a smaller screening radius for the same value of $r_{200}$, $r_{-2}$ and Chameleon parameters, resulting in tighter constraints in the space $(\phi_2,\mathcal{Q}_2)$, due to the additional un-screened region. Note that the same results have been found for clusters generated with other mass distributions besides the NFW and Burkert models.
    
    Results for case II are shown in Fig.~\ref{fig:case2} and \cref{tab:resMG}, where the two-$\sigma$ constraints on $\phi_2$ and $\mathcal{Q}_2$ is further reported. Note that in this notation, the fiducial values with which the halos are generated correspond to $\phi_2=0.65$ and $\mathcal{Q}_2=0.5$. Once again, no evident bias is found when adopting the ``wrong'' mass density model in the \textsc{MG-MAMPOSSt} fit. Overall, the fiducial values are always well within the $68\%$ confidence level limits, and the deviation from the GR case is evident in all the analyses performed, with $\mathcal{Q}_2$ generally better constrained than $\phi_2$.

    In both cases, it is worth noticing the difference in the marginalized contours between the halos generated assuming a Burkert profile (solid lines and filled regions in Figs.~\ref{fig:case1}, \ref{fig:case2}) and the one generated using a NFW model (dashed lines/contours). In particular, the NFW model generally provides wider constraints on the Chameleon parameters; as already mentioned above, this comes as a consequence of the effectiveness of the screening, which strongly depends on the assumed mass model, given the same background universe (GR or modified gravity).  

\begin{figure}
        \centering
        \includegraphics[width=0.9\linewidth]{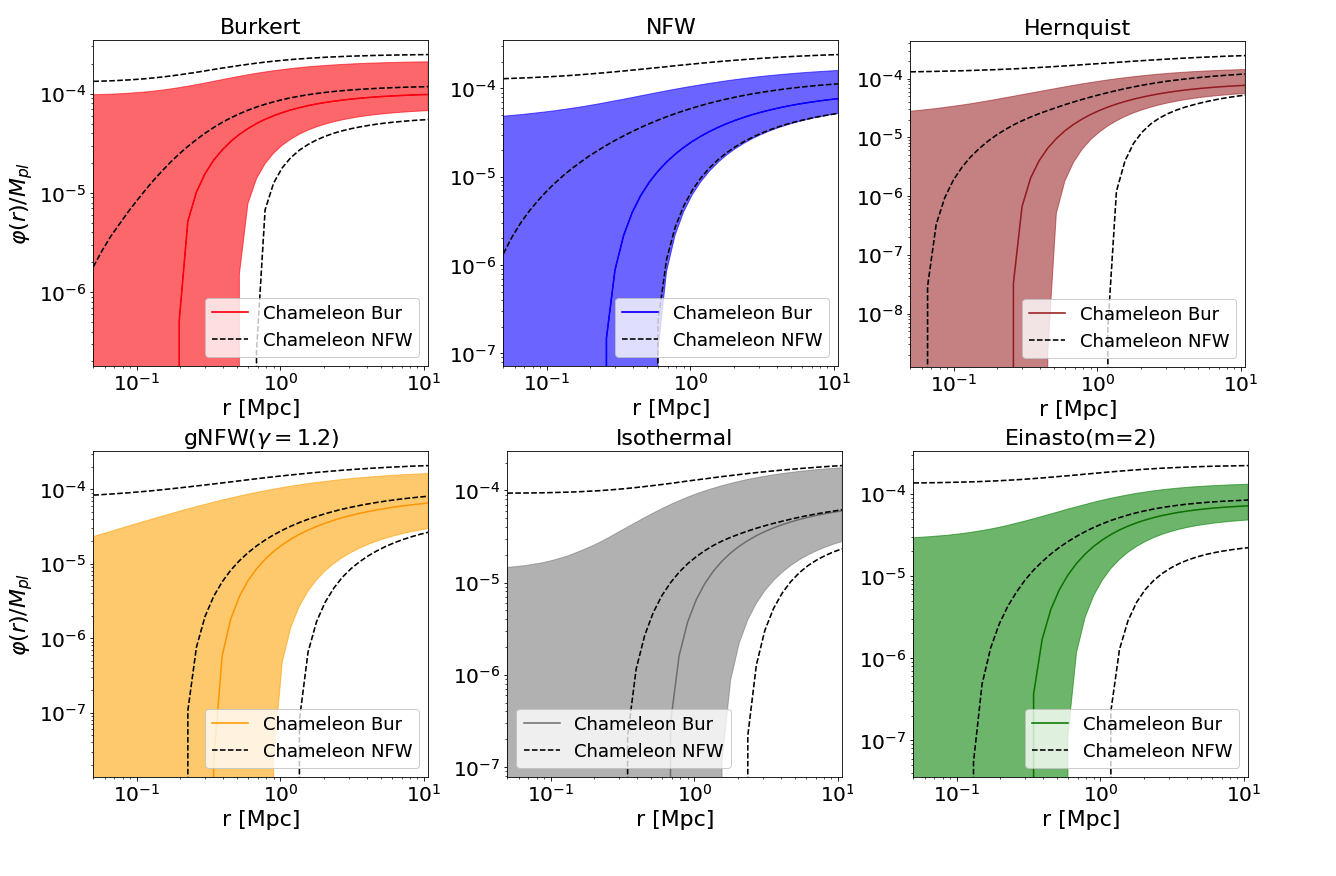}  
        \vspace{-0.2in}
        \caption{Radial field profiles $\phi/\Mp$, in unit of $c^2$, computed from the \textsc{MG-MAMPOSSt} posteriors of case II. Solid lines: Burkert-generated halo. Dashed line NFW-generated halo. The shaded areas (dashed top and lower lines) represent the $1\,\sigma$ regions.}
        \label{fig:fieldmarg}
    \end{figure}

    In Fig. \ref{fig:fieldmarg} a list of plots of the reconstructed field profiles for case II is further shown.  As expected, the stronger screening efficiency of the NFW model results in wider uncertainties in the field profile. For large radii, all profiles tend to the true background value $\phi/\Mp =10^{-4}c^2$, with some variability. It is worth mentioning that the best agreement is found when the true profile is used in the \textsc{MG-MAMPOSSt} analysis.
%
    \subsection{Bayesian evidence}
%
   {In order to determine possible strong statistical preferences among the mass models, the Bayesian evidence\footnote{To estimate the same we utilise \texttt{MCEvidence} \cite{Heavens:2017afc}, publicly available at {\href{https://github.com/yabebalFantaye/MCEvidence}{https://github.com/yabebalFantaye/MCEvidence}.}} $\logB$ \cite{Trotta:2017wnx, Trotta:2008qt} has been computed for each of the posteriors of the simulated phase space fitted against different mass models considered here. In the last column of \cref{tab:resMG}, the difference in the Bayesian evidence obtained using different mass models is shown. As anticipated, there exists no spurious preference for a varied mass model, which is a clear validation of the fact that the assumption of the mass model does not bias a detection or not of the fifth force. Recovering the posteriors of the Chameleon field to be consistent with the simulated clusters is in itself a validation of no bias. In addition, finding no strong deviation in terms of the Bayesian evidence validates our ability to constrain the same with future better observations of galaxy clusters.} 

   Note, however, that the assessment so far is done for case II when simulating the galaxy clusters in a strongly modified gravity regime. A similar comparison can be performed for the GR clusters simulated in case I. Even in this case, we find that the Bayesian evidence shows no bias for a spurious detection when an incorrect mass model is assumed in assessing the constraints on the Chameleon field. For the NFW-generated halo all the mass models assumed have $\logB < 0$, implying the NFW model is always preferred and a maximum disadvantage is found for the Burkert model with $\logB \sim -1.2$. When the halo is generated with the Burkert model, a slight preference ($\logB \sim 1.0$) is found for the Hernquist fitting model, but this is statistically not relevant.
    \begin{figure}
    \centering
     \includegraphics[scale=0.6]{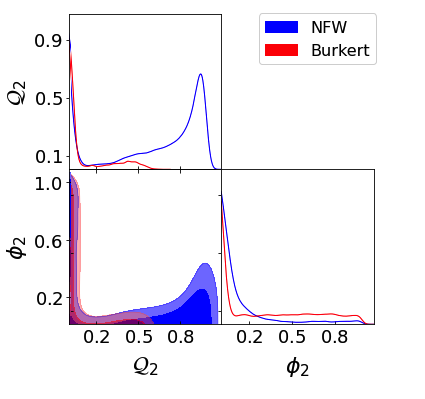}  
     \caption{Marginalized posteriors for the one and two-dimensional distributions of $\phi_2$ and $\mathcal{Q}_2$  from the \textsc{MG-MAMPOSSt} analysis of one synthetic NFW cluster with $10^4$ members within the virial radius. $r_{200}$ and $r_\text{s}$ are kept fixed in the fit. Blue: fit assuming an NFW ansatz. Red: fit assuming a Burkert ansantz.} 
     \label{fig:10000}
    \end{figure}

{\renewcommand{\arraystretch}{1.3}
\setlength{\tabcolsep}{4pt}
    \begin{table*}[]
     \centering
     \caption{2$\sigma$ constraints on the modified gravity parameters  $\phi_2 $,
      $\mathcal{Q}_2$ obtained from the \textsc{MG-MAMPOSSt} fit of two synthetic clusters generated in a Chameleon universe with $\phi_2 = 0.65$, $\mathcal{Q}_2 = 0.5$. The first six rows refer to a cluster whose true mass density follows a Burkert profile, and the last six rows correspond to a cluster generated assuming an NFW profile. Different mass density models are adopted in the fit, as shown in the second column. The last column shows the Bayesian evidence, $\logB$, for the varied models, contrasted against the reference model. }
        \begin{tabular}
         {c c | c c c }
         \hline
         \toprule
         \makecell{ Mass model \\ Simulation } & \makecell{ Mass model \\ Fitting }  & $\mathcal{Q}_2$ & $\phi_2 $ & $\logB$ \\
         \cline{1-5}
         \\
    Burket & Burket & $0.47^{+0.14 }_{-0.16}$ & $0.67^{+0.33 }_{-0.22}$ & 0.00 \\ [2mm]
    Burket & NFW & $0.46^{+0.16 }_{-0.17}$ & $0.65^{+0.35 }_{-0.20}$ & $+0.02$\\ [2mm]
    Burket & Hernquist & $0.50^{+0.15 }_{-0.15}$ & $0.60^{+0.40 }_{-0.20}$ & $+0.82$ \\[2mm]
    Burket & $g$NFW ($\gamma = 1.2$) & $0.51^{+0.11 }_{-0.11}$ & $0.58^{+0.42 }_{-0.32}$ & $-0.96$\\ [2mm]
    Burket & Isothermal & $0.57^{+0.14 }_{-0.14}$ & $0.57^{+0.41 }_{-0.28}$ & $+0.20$\\ [2mm]
    Burket & Einasto ($m=2$) & $0.52^{+0.15 }_{-0.15}$ & $0.57^{+0.43 }_{-0.22}$ & $+0.61$ \\ [2mm] 
    \cline{1-5}
    \\
    NFW & Burket & $0.49^{+0.18 }_{-0.16}$ & $0.68^{+0.32 }_{-0.31}$ & $-1.02$ \\ [2mm]
    NFW & NFW & $0.46^{+0.18 }_{-0.17}$ & $0.71^{+0.29 }_{-0.30}$ & 0.00 \\ [2mm]
    NFW & Hernquist & $0.48^{+0.18 }_{-0.16}$ & $0.73^{+0.27 }_{-0.41}$ & $+0.27$\\ [2mm]
    NFW & $g$NFW ($\gamma = 1.2$) & $0.47^{+0.12 }_{-0.12}$ & $0.62^{+0.38 }_{-0.45}$ & $-0.30$\\[2mm]
    NFW & Isothermal & $0.58^{+0.16 }_{-0.14}$ & $0.61^{+0.39 }_{-0.34}$ & $-0.82$\\ [2mm]
    NFW & Einasto ($m=2$) & $0.54^{+0.28 }_{-0.20}$ & $0.57^{+0.43 }_{-0.42}$ & $-0.63$\\ [2mm]
    \hline
    
         \bottomrule
        \end{tabular}
     \label{tab:resMG}
    \end{table*}
    }
%
\section{Discussions and conclusions}\label{sec:conclusions}
%
    This paper adopts a well-established procedure (see, \textit{e.g.} \cite{Terukina12}) to obtain semi-analytical solutions of the Chameleon scalar field profile assuming six matter density models describing the total mass distribution of galaxy clusters. The validity of the results is ensured by comparing the semi-analytic results with full numerical solutions of the Chameleon equation in the quasi-static limit.

    The work investigates the effect of the mass modeling in the kinematics analyses of clusters in Chameleon gravity by applying the \textsc{MG-MAMPOSSt} method to mock cluster-size halos, generated both in GR and in a strong modified gravity scenario. It has been found that possible systematic effects due to wrong mass modeling do not produce any noticeable impact on the distribution of the Chameleon parameters, resulting in reliable statistical uncertainties based on current imaging and spectroscopic surveys. If all the systematic effects are under control, the analysis of a GR-generated cluster always includes the GR limit ($\phi_\infty = 0,\, \beta = 0$) within the $1 \sigma$ region. Similarly, a Chameleon-generated halo exhibits a clear signature (for strong deviations from GR) of modified gravity for all the mass models used, with the obtained constraints in agreement with uncertainties.
    
    One can argue that the absence of biases due to the choice of the mass profile is just a consequence of the wide statistical uncertainties; as such, it is interesting to quantify the amount of systematic shift due to the wrong choice of mass model in an ideal situation where the statistical uncertainties are drastically reduced. A pps with more than $10^4$ particles within the virial radius following an NFW mass profile in GR was generated. A further perfect, unbiased, knowledge of the parameters $r_{200}$ and $r_\text{s}$ (\textit{i.e.} an infinitely precise prior, equivalent to set $\sigma_{r200},\sigma_{\rs} \to 0$ in the lensing Gaussian distribution)\footnote{One can think of this case as a sort of stacking of several clusters, even if in that case the effect of the mass bias would have been averaged out} was assumed. The ($\phi_2,\,\mathcal{Q}_2$) parameters are then constrained with the \textsc{MG-MAMPOSSt} code for the various mass density profile choices.

    Also, in this case, one finds consistency with GR and no evident systematic bias due to the modelling. This indicates that the mass profile reconstruction is robust against the choice of the matter density profile. The main difference is, once again, the shape of the constraints related to the efficiency of the Chameleon screening. As an example, Fig.~\ref{fig:10000} shows the marginalized distribution of ($\phi_2,\, \mathcal{Q}_2$) for the fit with an NFW model and a Burkert model (a similar trend is valid for the other profiles). 
    
    In conclusion, the effect of different mass distributions in the Chameleon screening mechanism is twofold: on one side, it sightly affects the constrained region of the parameter space, both for clusters generated in GR and in modified gravity. On the other hand, given the same decrease of the density at large radii, clusters in modified gravity with a cuspy density distribution (such as NFW) are characterized by a more efficient screening; thus, they could appear either completely screened (\text{i.e.} no signature of departures from GR), or they could give weaker constraints with respect to a more cored distribution (\text{e.g.} Burkert) even in strong background Chameleon universes. Profiles that decline as $r^{-\alpha}$ exhibit a smaller screening with increasing $\alpha$. This can be seen by a visual inspection of \cref{eq:sqbNFW}. From our analysis, it appears that the NFW ansatz has the most efficient screening among the models investigated.

    Note that, although the results have been presented here for a single halo size, it has been checked that possible biases are negligible when changing the parameters $r_{200},r_{-2}$ (or equivalently $\rhos,\rs$ of the mock halo). However, the efficiency of the screening mechanism strongly depends also on the halo size, as shown in \textit{e.g.} \cite{Pizzuti:2020tdl}; this affects the constrained region of ($\Qtwo,\phitwo$) in the joint kinematic+lensing analysis.  In particular, low mass halos with smaller $r_{200}$ and with a lower concentration $c_{200} = r_{200}/r_{-2}$ (i.e. small values of the product $r_\text{s}^2\rho_\text{s}$) are characterized by overall less screening. This can be also visually spotted by looking at the screening equations in Sec. \ref{sec:solutions}.
 
    As a completion, the bounds of Fig.~\ref{fig:10000} can be translated to bounds on the field $|f_{R,0}|$ in the popular $f(R)$ models of gravity \cite{buchadal70}, where Einstein-Hilbert action is modified by introducing a smooth function of the Ricci scalar $f(R)$, whose derivative $\text{d}f(R)/\text{d}R|_{z=0} = f_{R,0}$ behaves as a scalar field mediating the fifth force. The connection with the background Chameleon field is given by the mapping (see \textit{e.g.} \cite{Brax:2008}):
    \begin{equation}
        |f_R(z)| = \exp\left(-\frac{2\beta\phi_\infty(z)}{\Mp c^2}\right)-1\ .
    \end{equation}
    From the posterior of Fig.~\ref{fig:10000} one gets 
    $|f_{R,0}|\lesssim 4.4\times 10^{-6}$ for the NFW case and
    $|f_{R,0}|\lesssim 3.5\times 10^{-6}$ for the Burkert case, at $95\,\%$ C.L. upper limits. These bounds are in line with current cosmological constraints on the field (\textit{e.g.} \cite{Xu:2015,Wilcox:2015kna,Boumechta:2023qhd})\footnote{Note that tests at galactic scales \cite{Desmond2020} impose more stringent limits of $|f_{R,0}|\lesssim 10^{-8}$, when a particular model for the $f(R)$ function is adopted.}, and provide the limiting precision that can be achieved by kinematics and lensing analysis of clusters. It is worth to point out that the NFW model exhibits a bump in the allowed region at large $\Qtwo$, which is a consequence of the stronger efficiency of the screening mechanism. Indeed, as one can visually inspect form \cref{eq:NFWscreen}, increasing the coupling $\beta$ results in an increase of the screening radius, suppressing the fifth force for reasonably small $\phitwo$ (see also the discussion in \cite{Pizzuti:2020tdl}).
    
    Note that the analysis of \cite{Tamosiunas_2022} on $\Lambda$CDM cosmological simulation demonstrates that the Chameleon-to-GR acceleration ratio at $r_{200}$, given current bounds on the chameleon parameters, should not be greater than $10^{-7}$. This is well below the uncertainties on any cluster's mass profile reconstruction. The results have been shown for the NFW case truncated at the virial radius, which exhibits a stronger screening effect compared to other mass models. Moreover, the current mass determination can be extended above the virial radius (\textit{e.g.} \cite{Butt:2024jes}), but also, in this case, one expects the effect of the Chameleon acceleration to be hardly detectable.
    
    However, it is still interesting to investigate the possible imprint that the fifth force can leave in clusters when the different mass components are accurately modelled. In \cite{Sartoris2020, Biviano:2023oyf} showed that kinematics data of galaxies in clusters and the stellar velocity dispersion of the Brighter Cluster Galaxy (BCG) can be combined to obtain a very accurate multi-component mass determination of clusters down to the very centre ($\lesssim 1\,\text{kpc}$). A detailed modelling of the screening mechanism for each matter distribution over three orders of magnitude in scales could, in principle, produce peculiar signatures of the fifth force, which will be investigated in an upcoming work.
    
    We also point out that the effect of halo triaxiality has not been taken into account. As shown by \cite{Tamosiunas_2022}, this can enhance the effect of the fifth force up to $50\%$ at the virial radius. Moreover, the simulation implemented with \textsc{ClusterGEN} assumes that halos are isolated; in reality, clusters are connected to the large-scale structure, and the environmental screening can play a significant role (see \textit{e.g. }\cite{Burrage:2017shh}). To model correctly the screening as well as to determine which could be the best mass density profile to describe halos in a Chameleon universe, it would be interesting to apply the semi-analytical solutions found here to cosmological simulation in modified gravity. This is, however, beyond the scope of this paper. 

    {Such an investigation as performed here, is very crucial also in light of current and upcoming surveys, such as NIKA \cite{Ruppin:2017bnt} or EUCLID \cite{Euclid:2019bue}, which will provide a large amount of data down to an unprecedented level of precision. The analysis performed here serves as a solid base to carry out joint analyses of relaxed galaxy clusters, combining information from lensing, kinematics of galaxies and the hot gas of the Intra Cluster Medium. Another important extension is to model the chameleon field coupling to different mass components' contribution to the total mass of the galaxy cluster. As the future observations of the galaxy clusters become more resolved it is essential to move away from the simple single mass profile coupled to chameleon field modelling, which is underway. }

\section*{Acknowledgements}
We thank the anonymous Referee for her/his precious comments and suggestions. BSH is supported by the INFN INDARK grant and acknowledges support from the COSMOS project of the Italian Space Agency (cosmosnet.it). CB acknowledges support from the COSMOS project of the Italian Space Agency (cosmosnet.it), and the INDARK Initiative of the INFN (web.infn.it /CSN4 /IS /Linea5/ InDark). A. M. Pombo is supported by the Czech Grant Agency (GA\^CR) under grant number 21-16583M. AL has been supported by the EU H2020-MSCA-ITN-2019 Project 860744 `BiD4BESt: Big Data applications for Black Hole Evolution Studies' and by the PRIN MIUR 2017 prot. 20173ML3WW, `Opening the ALMA window on the cosmic evolution of gas, stars, and supermassive black holes'.
    
\bibliographystyle{JHEP}
\bibliography{bibliography}

\end{document}